\documentclass[12pt,preprint]{aastex}

\usepackage{amsmath}
\usepackage{graphicx}
\usepackage{multirow}
\usepackage{color}
\usepackage{natbib}
\usepackage{wasysym}

\graphicspath{{Figures/}}

\font\sm=cmss10 at 10pt

\begin{document} 

\title{The current impact flux on Mars and its seasonal variation}

\author{Youngmin JeongAhn and Renu Malhotra}
\affil{Lunar and Planetary Laboratory, The University of Arizona, Tucson, AZ 85721, USA.}
\email{jeongahn@lpl.arizona.edu,renu@lpl.arizona.edu}

%\slugcomment{\today}

\begin{abstract}
We calculate the present-day impact flux on Mars and its variation over the Martian year, using the current data on the orbital distribution of known Mars-crossing minor planets.  We adapt the {\"O}pik-Wetherill formulation for calculating collision probabilities, paying careful attention to the non-uniform distribution of the perihelion longitude and the argument of perihelion owed to secular planetary perturbations.  We find that these previously neglected non-uniformities have a significant effect on the mean annual impact flux as well as its seasonal variation. The impact flux peaks when Mars is at aphelion, but the near-alignment of Mars' eccentricity vector with the mean direction of the eccentricity vectors of Mars-crossers causes the mean annual impact flux as well as the amplitude of the seasonal variation to be significantly lower than the estimate based on a uniform random distribution of perihelion longitudes of Mars-crossers.  We estimate that the flux of large impactors (of absolute magnitude $H<16$) within $\pm30^\circ$ of Mars' aphelion is about four times larger than when the planet is near perihelion.  Extrapolation of our results to a model population of meter-size Mars-crossers shows that if these small impactors have a uniform distribution of their angular elements, then their aphelion-to-perihelion impact flux ratio would be as large as 25.  These theoretical predictions can be tested with observational data of contemporary impacts that is becoming available from spacecraft currently in orbit about Mars. 
\end{abstract}

\section{Introduction}\label{s:intro}
The impact crater record on the surfaces of the terrestrial planets over geologically long timescales provides a window on the dynamical history of the solar system, including a chronology of major geological and dynamical events.  This crater-based chronology is calibrated primarily on estimates of the cratering rate on the Moon over geological timescales.  The presence of space observatories in orbit about Mars now offers a new opportunity to measure the present-day impact flux on Mars in statistically significant numbers, at least for small impactors, and thereby improve the calibration of crater-based chronology for geologically recent times.  Within the last $\sim$~decade, more than 200 dated impact sites have been identified, by comparison of before- and after- images of Mars' surface, including some that can be narrowed down to within a Martian year~\citep{Malin:2006,Daubar:2013}.  Also over the past $\sim$~decade, large scale and deep surveys for near-Earth objects have dramatically increased our knowledge of the population of minor planets in the inner solar system, hence the size and orbital distribution of the current Martian impactor population can be identified.  Thus, it is valuable to investigate the theoretically expected Martian impact rate based on the known impactor population, and to compare it with observations of contemporary Martian impacts.

The impact rate on Mars has been the subject of several previous studies.  
\citet{Ivanov:2001}  adapted the lunar impact crater production function 
to calculate the long-term impact crater production rate on Mars.  A series of studies \citep[and references therein]{Hartmann:2001,Hartmann:2005,Chapman:2004,Hartmann:2010}  have calculated increasingly refined  estimates of Martian cratering, taking account of complexities associated with secondary cratering and other effects.  \citet{LeFeuvre:2008} and \citet{LeFeuvre:2011} 
used a model of the orbital distribution of planet-crossing asteroids and comets to calculate the Martian cratering rate.  \cite{Williams:2014} 
used the observed annual flux of terrestrial fireballs to calculate the present-day production rate of small Martian craters.  In many of these studies, the Mars/Moon ratio of impact cratering is a fundamental parameter, and the studies attempt to obtain the impact crater production rate averaged over geological timescales for use in crater-based chronologies.  

In the present paper, we make an independent and improved estimate of the current impact flux on Mars, based on the known observational data set of the present-day population of Mars crossing objects (MCOs).  We demonstrate that the MCOs have an intrinsically non-random distribution of the longitude of perihelion $\varpi$ and argument of perihelion $\omega$, owing to secular planetary perturbations.  Such non-uniformity has been noted previously: the non-uniform $\omega$ circulation cycle of MCOs was briefly discussed as a protection mechanism against collisions on Mars \citep{Wetherill:1974}.  The  concentration of $\varpi$ in the asteroid belt was investigated for its role in the impact flux amongst main belt asteroids and Hildas \citep{Dell'Oro:1998,Dell'Oro:2001}.  However, detailed study of the non-uniform distribution of the angular elements of MCOs has not been carried out, nor its effects on the impact rate on Mars.
Therefore, we first identify the non-uniformities in the distribution of $\varpi$ and $\omega$ in the unbiased population of bright MCOs, then we develop a numerical method for computing the impact probabilities that takes account of these non-uniformities.  
We demonstrate that these non-uniformities have a significant effect on both the mean impact flux as well as its variation over Mars' seasons.   
Our calculations are based on a modification of the method of \citet{Wetherill:1967} and \citet{Opik:1951} which computes the probability of collision of two objects that move on fixed, independent Keplerian orbits.  The modification that we introduce accounts directly for the non-randomness of the angular elements of MCOs.  We also account in detail for gravitational focusing effects; these are also significantly affected by the non-random distribution of the angular elements of MCOs because the encounter velocities of MCOs with Mars are correlated with their angular elements. 

The rest of this paper is organized as follows.  In Section~2, we describe the observational data set of MCOs, derive an observationally complete subset, and describe the dynamically significant aspects of the distribution of its orbital elements.  In Section~3, we describe the details of our impact flux calculations and present our results.  In Section~4, we provide a comparison of our results with the results of previous studies, and in Section~5, we summarize and conclude.

\section{Mars Impactor Population}\label{s:pop}

\subsection{Definitions and Dynamical Considerations}\label{s:defns}

To estimate the current impact flux on Mars, we need to know the intrinsic orbital distribution of current Mars Crossing objects (MCOs), free from observational selection effects.  The orbital distribution is defined on the parameter space of the five orbital elements,  semimajor axis, eccentricity, inclination, argument of perihelion and longitude of perihelion, $(a,e,i,\omega,\varpi)$.  We obtain this by first looking to the current data on MCOs, and using a brightness cutoff to define a subset that is less affected by observational bias; we also estimate a correction factor for the observational completeness of this subset. We adopt the orbital distribution of the bright MCOs as a proxy for the orbital distribution of the entire population of Mars impactors.  It is conceivable that the orbital distribution of the population of very small MCOs differs from that of the bright (large) MCOs; we discuss this point in Section~\ref{s:smallimpactors}.

The current population of MCOs can be defined with cutoffs in perihelion and aphelion distances, analogous to  the common definition of Near Earth Objects (NEOs). Thus, the aphelion of MCOs should be greater than the current perihelion distance of Mars, 1.38~AU, and their perihelion should be smaller than the current aphelion of Mars, 1.67~AU. 
We obtained the current data on all minor planets from the Minor Planet Center (MPC), which maintains up-to-date lists of the minor planets of the solar system.  The dataset ``MPCORB" at epoch 15 June 2014 contains the osculating orbital elements as well as the absolute brightness magnitude of more than $6\times10^5$ minor planets. (We do not include objects that have poorly determined orbit solutions from only a single opposition of observation.)  Of these, only 16702 objects meet the definition of MCOs.

It is of some importance to note that the orbital parameters of Mars (and of the planet-crossing population of minor planets) vary significantly over geological timescales. 
On megayear timescales, Mars' orbital eccentricity varies from near zero to 0.12, and its orbital inclination also changes from near zero to 7 degrees relative to the invariable plane of the Solar system~\citep{Ito:2002}.  The current eccentricity, 0.093, is close to its long term maximum value, whereas the current inclination, $1.85$ degrees to the ecliptic ($1.67$ degrees to the invariable plane), is closer to its minimum value. The current semimajor axis of Mars is 1.524~AU; over megayear timescales, the semimajor axis variation is only 0.02\% from maximum to minimum (determined from our own numerical integration for a timespan of 100 myr).   Mars' perihelion longitude, $\varpi$, is currently at $336^\circ$ and its longitude of ascending node, $\Omega$, is at $50^\circ$, with respect to the ecliptic and equinox of J2000, and their current rates of change are $0.43$ and $-0.28$ degrees per century, respectively~\citep{Murray:1999}.  

Most previous studies have used estimates of the long-term average value of the impact rate to draw isochrones \citep[e.g.,][]{Williams:2014}.  This rate is time-variable as Mars' eccentricity varies on secular timescales, therefore, the current value of Mars' eccentricity, 0.093, should be used when the current impact rate is of interest.  We are focusing on the current impact rate on Mars in this paper but the sensitive nature of the impact rate with respect to time-varying Martian eccentricity will be briefly discussed later, in Section~\ref{s:comparison}.

Martian climate seasons are conventionally defined by its vernal and autumnal equinoxes and according to the solar longitude, $L_S$, in reference to its obliquity. However, as we will see below, Mars' variable heliocentric distance is a major factor that determines the annual variation of the impact flux.  Therefore, the ``seasonal variation'' of the impact flux on Mars is logically defined with respect to Mars' perihelion longitude, which corresponds to $L_S = 251^\circ$.  The northern hemisphere Winter roughly corresponds to the time of perihelion of Mars.

%%%%%%%%%%%%%%%%%%%%%%%%%%%%%%%%%%%%%%%%%%%%%%%%%%%%%%%%%%%%%%%%%%%%%%%%%
\subsection{Orbital distribution of MCOs}\label{s:orbits}

%%%%%%%%%%%%%%%%%%%%%%%%%%%%%%%%%%%%%%%%%%%%%%%%%%%%%%%%%%%%%%%%%%%%%%%%%
\begin{figure}
\centering
     \includegraphics[scale=0.1]{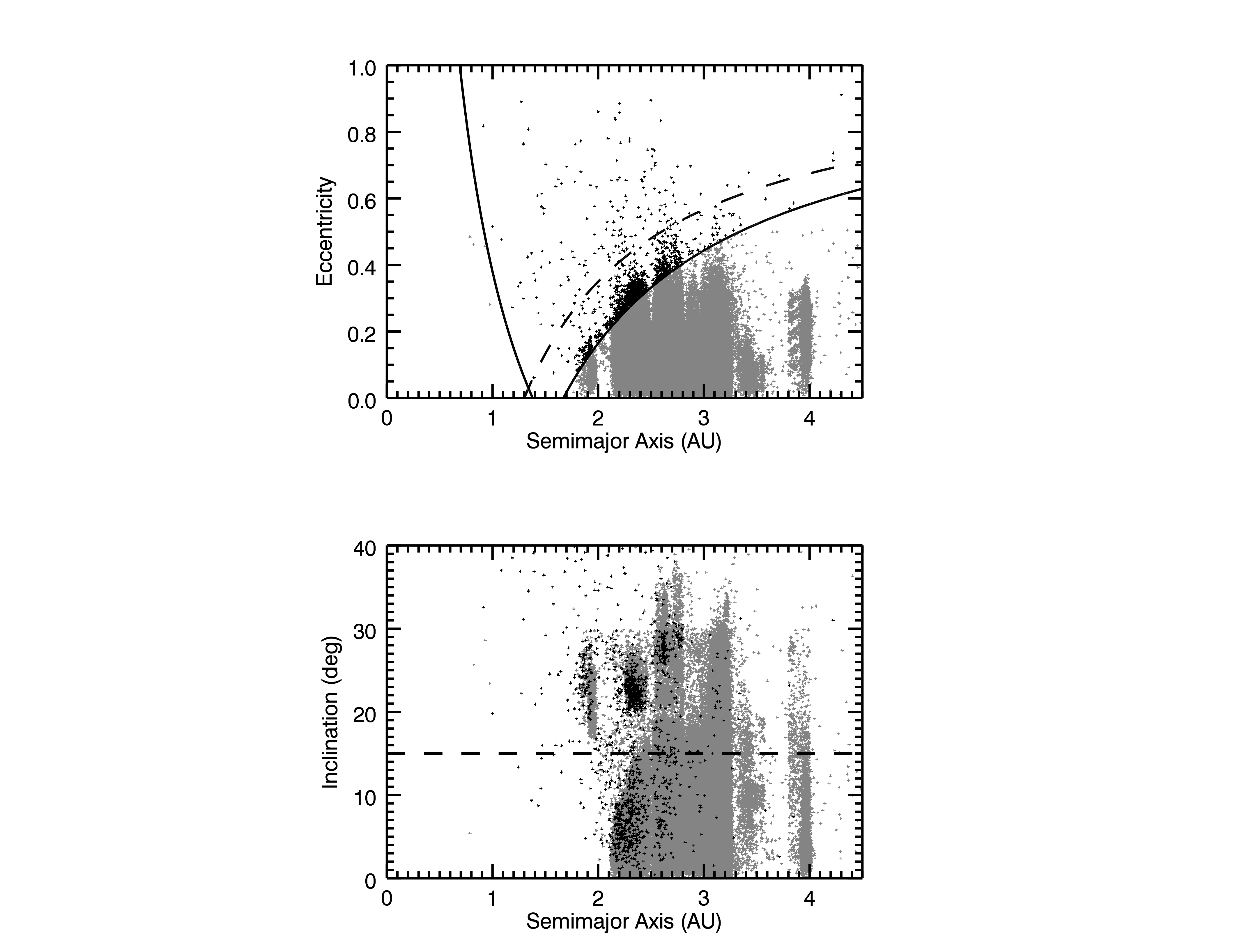}
  \caption{\sm Orbital distribution of bright ($H<16$) minor planets in the inner solar system.  The horizontal axes denote semimajor axes and the vertical axes are eccentricity and inclination in the upper and lower panel, respectively. Mars crossers are marked as black points, all others are in gray. The continuous line curves in the upper panel define the inner and outer limits of Mars crossers; the dashed curve indicates the outer boundary of Amors at perihelion distance $q=1.3$~AU. In the lower panel, the boundary of the ``high inclination'' and ``low inclination'' subgroups is indicated with the horizontal dashed line at $i = 15^\circ$. }
\label{AEI}
\end{figure}
%%%%%%%%%%%%%%%%%%%%%%%%%%%%%%%%%%%%%%%%%%%%%%%%%%%%%%%%%%%%%%%%%%%%%%%%%

A scatter plot of the semimajor axes and eccentricities of bright MCOs, of absolute magnitude $H<16$, is shown in the upper panel of Fig.~\ref{AEI}.  Most of this sample (about 85\% of 1322 objects) has semimajor axes in the range between 2.1~AU and 3.3~AU; note that this range overlaps with the main asteroid belt.  In this sample, 167 (13\%) objects have perihelion distance smaller than 1.3~AU and therefore also belong to the population of minor planets defined as near-Earth objects. The number of bright MCOs, $H<16$, that actually cross Earth's orbit (perihelion distance $q<1$~AU) is 84 (6\%).  The small body database of Jet Propulsion Laboratory (JPL) excludes NEOs from MCOs, for classification purposes; this was also done in a classification study of MCOs, \citet{Michel:2000}. However, in the work we carry out here,  we do not discard the current Earth-crossing population from MCOs because this population also contributes to the current Martian impact statistics.

A scatter plot of the semimajor axes, $a$, and inclinations, $i$, is shown in the lower panel of Fig.~\ref{AEI}.   In this plot, a clear division between ``low inclination'' and ``high inclination'' subsets of MCOs is evident, with a boundary at $i=15^\circ$.  A finer set of divisions was defined by~\citet{Michel:2000} who classified the MCOs into five different categories based on the three orbital parameters ($a,e,i$).   
According to the definitions of~\citet{Michel:2000}, the dividing line of $i=15^\circ$ corresponds to the lower inclination boundary of the Hungaria group of asteroids and it roughly matches with the location of the $\nu_6$ secular resonance which separates the Phocaea asteroid group and the highly inclined outer main belt asteroids (MB2, in their notation) from the main belt asteroids of low inclination (MB, in their notation). 
Here we distinguish only the two broader groups, ``low inclination'' and ``high inclination'' MCOs, with the boundary at $i=15^\circ$. As we will show, these two groups differ systematically in the distributions of their angular elements as well as in their impact velocities and their probabilities of collision with Mars.

%%%%%%%%%%%%%%%%%%%%%%%%%%%%%%%%%%%%%%%%%%%%%%%%%%%%%%%%%%%%%%%%%%%%%%%%%
\subsection{Absolute magnitude distribution and completeness of bright MCOs}\label{s:Hdist} 

%%%%%%%%%%%%%%%%%%%%%%%%%%%%%%%%%%%%%%%%%%%%%%%%%%%%%%%%%%%%%%%%%%%%%%%%%
\begin{figure}
\centering
  \includegraphics[scale=0.7]{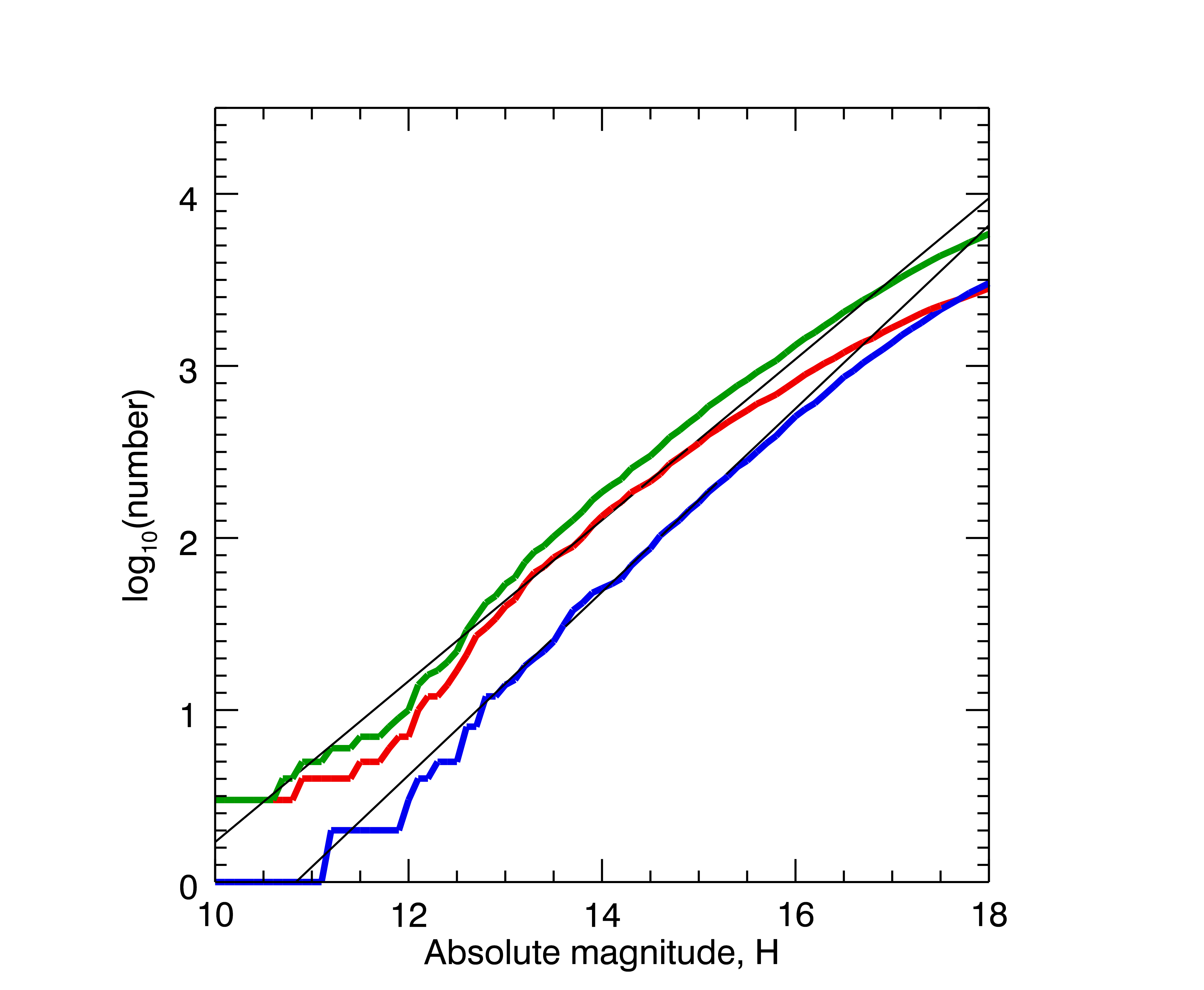}
  \caption{\sm Cumulative distribution of the absolute magnitude, $H$, of known Mars crossing objects.  Note that the ordinate is on a logarithmic scale.  The green curve indicates all MCOs,  the red and blue curves indicate the high inclination $(i>15^\circ$) and the low inclination $(i<15^\circ$) subsets.  The two black diagonal lines indicate the best-fit linear functions for high and low inclination MCOs with $13<H <15$. }
\label{hmag}
\end{figure}
%%%%%%%%%%%%%%%%%%%%%%%%%%%%%%%%%%%%%%%%%%%%%%%%%%%%%%%%%%%%%%%%%%%%%%%%%

The absolute brightness magnitude of MCOs ranges from $H=9.4$ (132 Aethra) to $H=33.2$ (2008 TS26), however the population of fainter objects is highly incomplete due to many observational selection effects.  Figure~\ref{hmag} plots the cumulative $H$ magnitude distribution of MCOs up to $H=18$.  The MPC database lists $H$ magnitudes with only one decimal place accuracy; for the binned distribution,  the MCOs corresponding to boundary values were allotted equally to adjacent bins.  We observe that the slope of the $H$ distribution decreases noticeably  for $H>15$, indicating increasing observational incompleteness; therefore, we assume that MCOs are observationally complete for $H<15$.  However, there are only 516 MCOs with $H<15$.  To obtain better statistics for their orbital distribution, we adopted the sample of MCOs having $H<16$.  This sample consists of 1322 objects (8\% of the 16702 known MCOs with well-determined orbits); hereafter, we refer to this set as the ``bright MCOs''.  

To calculate the incompleteness level of our sample of bright MCOs, we estimated their intrinsic number as follows.   We observe in Fig.~\ref{hmag}, that both the low (blue curve) and high (red curve) inclination MCOs follow an almost straight line, hence a single power law distribution, in the range $13< H < 15$.   We calculated the least-squares  best-fit linear function for the cumulative distribution of MCOs in this brightness range.  The best-fit slope of the low inclination population, $0.533\pm 0.009$, is found to be slightly higher than that of the high inclination population, $0.468\pm 0.007$.   (We quote 1--$\sigma$ uncertainties of the best-fit slopes.)   Because the difference in slopes of the low and high inclination groups is more than 7--$\sigma$, we assume that it is real. 
(We also calculated the best-fit linear function for the smaller  range, $13<H<14$.   The slopes of both populations are steeper in this range compared to the slopes over the 13--15 $H$-magnitude range, 
but again the low inclination MCOs have a steeper slope, $0.603\pm 0.020$, than the high inclination ones, $0.504\pm 0.018$.  The cause of the steeper slope within $H$ of 13--14 compared to $H$ within 13--15 could be the wavy character of the size and brightness distribution, but undetected objects near $H$ of 15 can also contribute to this trend.)  
A steeper slope at low inclinations was also found for main belt asteroids~\citep{Terai:2011}. 

With the above best-fits, we can describe the cumulative number distribution of MCOs of brightness  $H\gtrsim13$ as
%%%%%%%%%%%%%%%%%%%%%%%%%%%%%%%%%%%%%%%%%%%%%%%%%%%%%%%%%%%%%%%%%%%%%%%%%
\begin{equation}
N(<H) = A_L \times 10^{\alpha_L (H-15)} + A_H \times 10^{\alpha_H (H-15)},
\label{cumdist}
\end{equation}
%%%%%%%%%%%%%%%%%%%%%%%%%%%%%%%%%%%%%%%%%%%%%%%%%%%%%%%%%%%%%%%%%%%%%%%%%
where the first term, with $A_L=166\pm4$ and $\alpha_L=0.533\pm 0.009$, describes the low inclination MCOs, and the second term, with $A_H=373\pm7$ and $\alpha_H=0.468\pm 0.007$, describes the high inclination MCOs. 

With this approximation, we estimate that the total number of bright MCOs ($H<16$) is 1662, and the observed sample of 1322 represents 80\% of this intrinsic population. Therefore, the undetected MCOs ($H<16$) are about 11\% and 35\% of the observed MCOs, in the low inclination and the high inclination subgroups, respectively.

We note that we have implicitly  assumed that the $H$ distribution of bright MCOs is largely independent of their orbital distribution (save for the inclinations, as discussed). However, the $H$ magnitude distribution may vary in orbital parameter space due to size-dependent local collisional history or non-gravitational effects such as the thermal Yarkovsky  drift. The different power law distributions of the high and low inclination subgroups of MCOs we calculated above reflect part of this complex reality.  Another example of this complexity is the V-shaped dispersion of the Hungaria family in the $a-H$ diagram due to Yarkovsky drift \citep{Warner:2009}.

We understand also that the Hungarias have higher average albedos than  other MCOs due to a high fraction of E-type asteroids \citep{CanadaAssandri:2015}. Therefore, by choosing a single $H$ magnitude cutoff, our sample includes smaller size Hungarias than in the other populations. However, the number of Hungarias in our sample is 113, accounting for only 8.5\% of the bright MCOs; therefore we keep using the single magnitude cutoff, $H<16$.

%%%%%%%%%%%%%%%%%%%%%%%%%%%%%%%%%%%%%%%%%%%%%%%%%%%%%%%%%%%%%%%%%%%%%%%%%
\subsection{Non-uniform Apsidal Distribution}

%%%%%%%%%%%%%%%%%%%%%%%%%%%%%%%%%%%%%%%%%%%%%%%%%%%%%%%%%%%%%%%%%%%%%%%%%
\begin{figure}
\centering
  \includegraphics[width=300px]{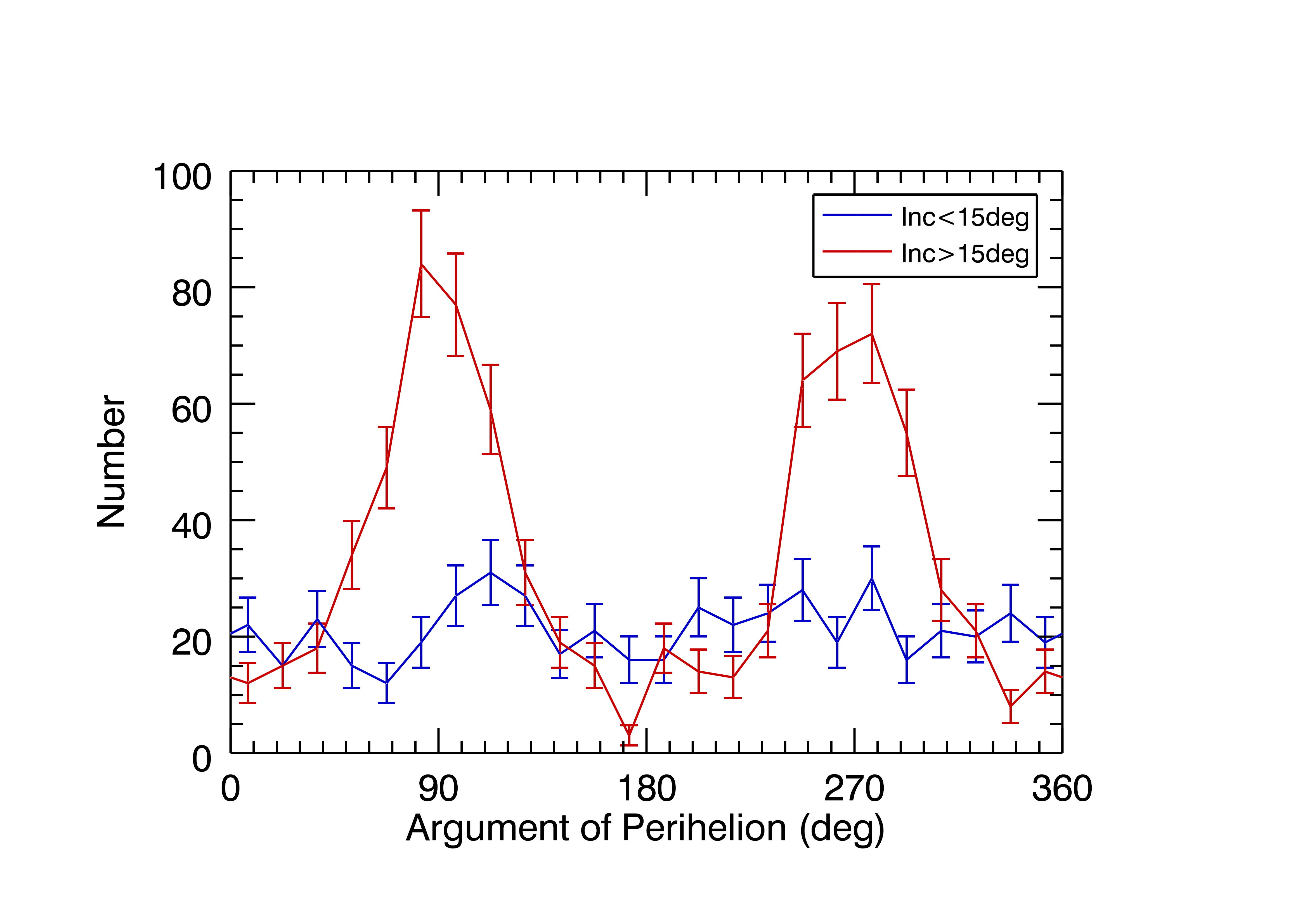}
    \caption{\sm The distribution of the argument of perihelion, $\omega$, of bright MCOs ($H<16$), in two different inclination regimes.  }
\label{AGP2}
\end{figure}
%%%%%%%%%%%%%%%%%%%%%%%%%%%%%%%%%%%%%%%%%%%%%%%%%%%%%%%%%%%%%%%%%%%%%%%%%

The distribution of the argument of perihelion, $\omega$, of bright MCOs is shown in Fig.~\ref{AGP2}.  The MCOs' $\omega$ distribution is not independent of their inclinations. For the high inclination MCOs ($i>15^\circ$, sample size $N=813$ objects) the concentrations around 90 and 270 degrees are evident. One possible reason for this concentration is the $i-e-\omega$ coupling caused by Jupiter's perturbation~\citep{Kozai:1962}.   This coupling, which occurs preferentially for high inclination orbits, causes the eccentricity to reach a maximum when $\omega$ is near $90^\circ$ or $270^\circ$.  The maxima in eccentricity enable many of the high inclination objects to just barely be classified as MCOs.  Thus, the boundary of MCOs in parameter space selectively contributes to the $\omega$ concentration. Secondly, long residence times around $\omega \simeq 90^\circ$ and $270^\circ$ (due to the slow rate of change of $\omega$ near these values) may also contribute to the concentrations.

%%%%%%%%%%%%%%%%%%%%%%%%%%%%%%%%%%%%%%%%%%%%%%%%%%%%%%%%%%%%%%%%%%%%%%%%%
\begin{figure}
\plottwo{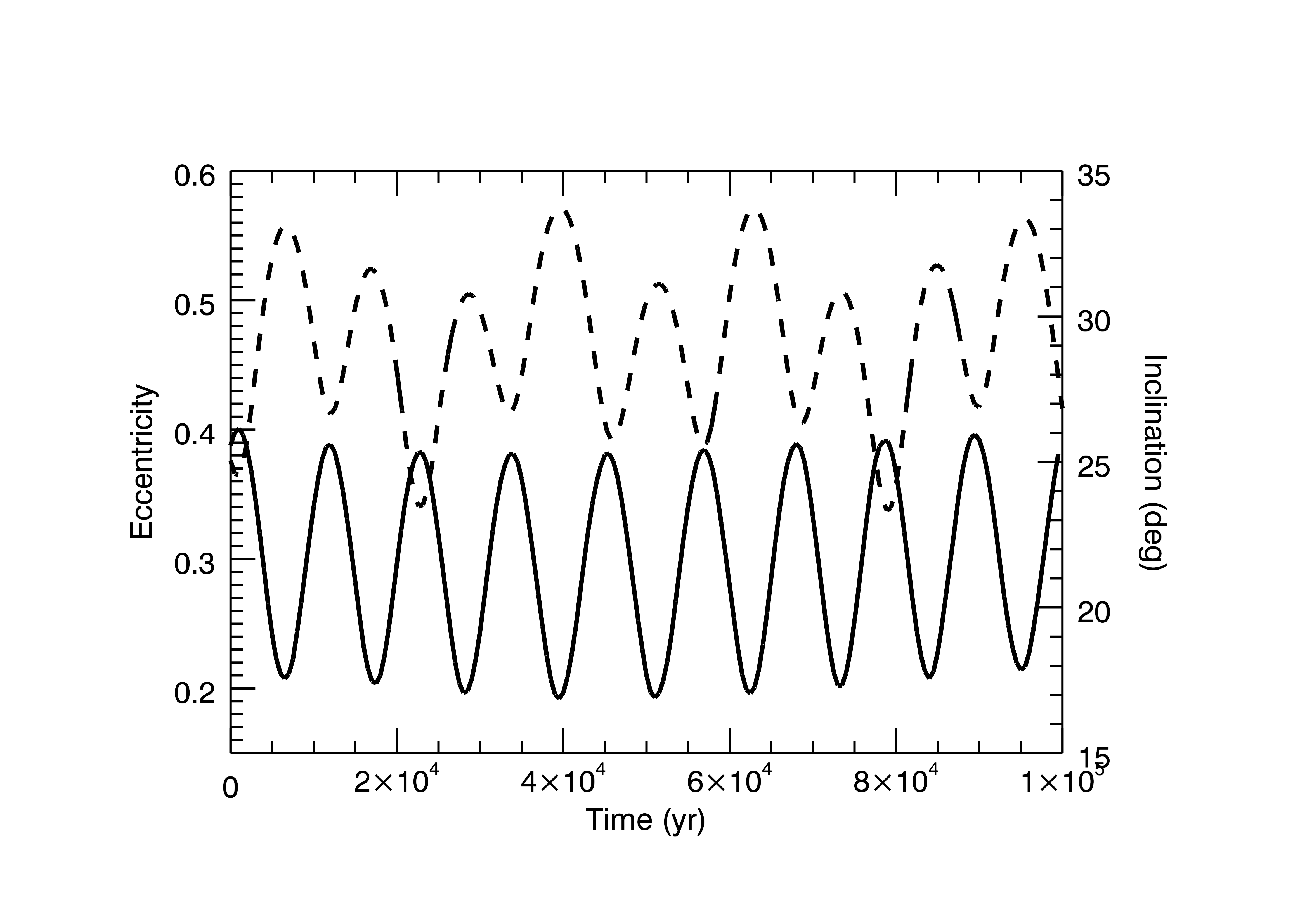}{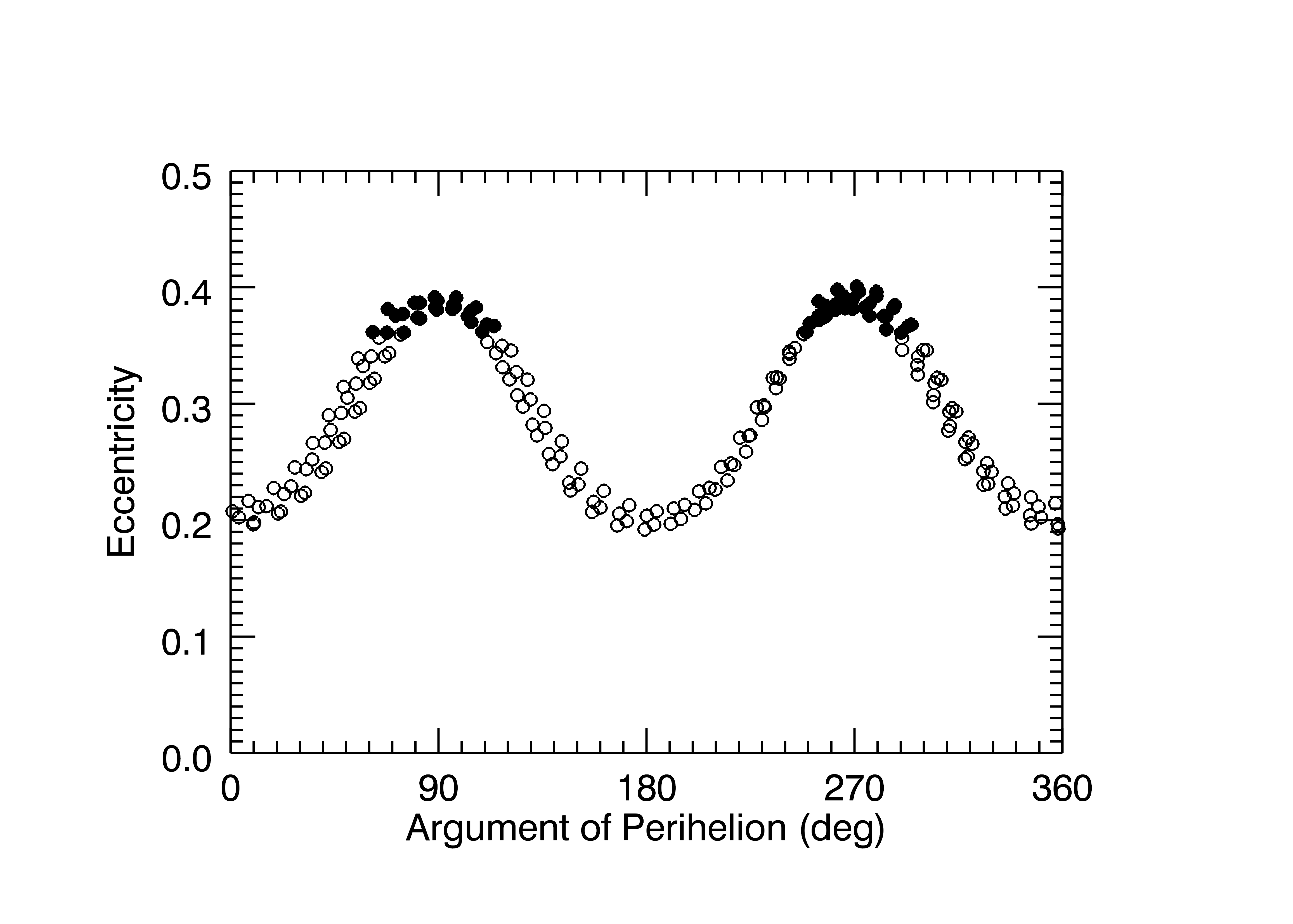}
  \caption{\sm The orbital evolution of 132 Aethra over $10^5$ years, including the gravitational perturbations from the planets.  (a) The evolution of eccentricity and inclination. (b) The coupling between eccentricity and argument of perihelion; the filled circles indicate the times when Aethra's perihelion distance is less than Mars' aphelion distance.  }
\label{f:Aethra}
\end{figure}
%%%%%%%%%%%%%%%%%%%%%%%%%%%%%%%%%%%%%%%%%%%%%%%%%%%%%%%%%%%%%%%%%%%%%%%%%

The coupled oscillations of $i-e-\omega$ are illustrated in the orbital evolution of 132 Aethra, the brightest and possibly the first discovered Mars' crossing asteroid.  Its current orbital elements, $(a,e,i)=(2.608,0.389,25^\circ)$, are rather typical of the high inclination bright MCOs.  We integrated 132 Aethra's orbit over a period of $10^5$ years, including the gravitational perturbations of all eight planets. We found that its semimajor axis variation is about 0.3\%, but its eccentricity and inclination exhibit large amplitude variations of opposite phase, with a period of about $10^4$ years, as shown in Fig.~\ref{f:Aethra}a.  The large eccentricity of Aethra at the present epoch makes it a Mars-crosser, but lower eccentricity over time changes this circumstance.  The maxima in the eccentricity occur at $\omega=90^\circ$ and $\omega=270^\circ$, as shown in Fig.~\ref{f:Aethra}b. In this figure, we also highlight the range when Aethra's perihelion distance is less than Mars' aphelion distance; we observe that this occurs when its argument of perihelion is near $\omega=90^\circ$ and $\omega=270^\circ$.   Further study is needed to confirm that this is the dominant behavior amongst high inclination MCOs. 

In some contrast with the high inclination subset of MCOs, the low inclination subset  ($i<15^\circ$, sample size $N=509$) does not show as clear an axial distribution of $\omega$.  We calculate the mean direction of $\omega$ and its significance level based on circular statistics for axial distributions \citep{Fisher:1993}.  We find that the mean direction is along $\bar{\omega}= 101^\circ$ and $281^\circ$; however, the Rayleigh test shows that the non-uniformity is not statistically significant; the probability $p$ that the distribution is uniform on the circle is greater than 5\%.  Therefore, for the currently available observational data set, we conclude that the low inclination MCOs are uniformly distributed in argument of perihelion.

%%%%%%%%%%%%%%%%%%%%%%%%%%%%%%%%%%%%%%%%%%%%%%%%%%%%%%%%%%%%%%%%%%%%%%%%%
\begin{figure}[h]
\centering
  \includegraphics[width=300px]{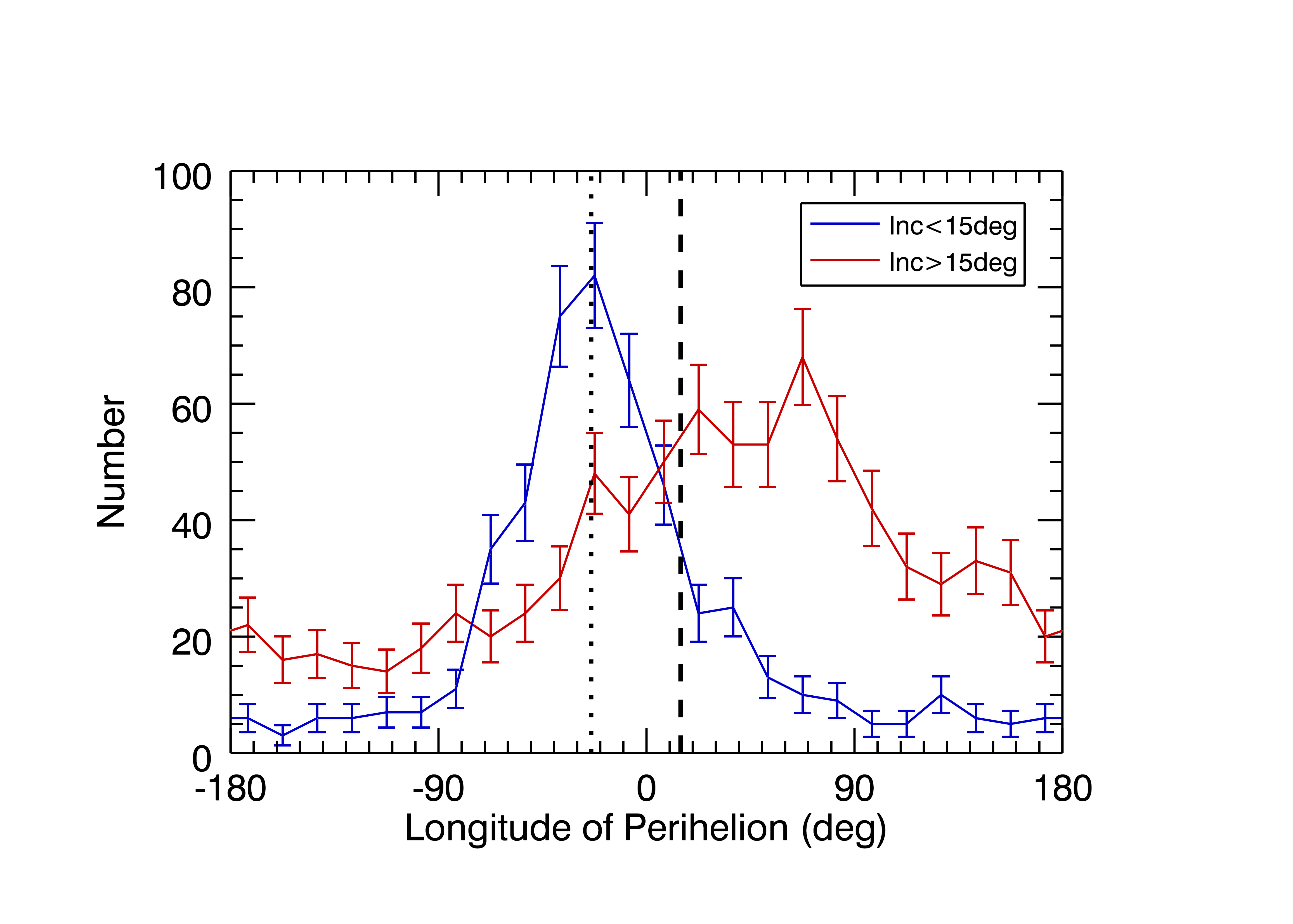}
  \caption{\sm The distribution of the longitude of perihelion, $\varpi$, of bright MCOs ($H<16$), in two different inclination regimes. The current value of the longitude of perihelion of Mars, $\varpi_{\mars}$, and of Jupiter, $\varpi_{\jupiter}$, are indicated by the vertical dotted and dashed lines, respectively.  }
\label{LNP2}
\end{figure}
%%%%%%%%%%%%%%%%%%%%%%%%%%%%%%%%%%%%%%%%%%%%%%%%%%%%%%%%%%%%%%%%%%%%%%%%%

The longitude of perihelion, $\varpi$, of bright MCOs has a strong unidirectional concentration, as seen in Fig.~\ref{LNP2}. This intrinsic concentration is also caused by secular effects due to planetary perturbations.  For main belt asteroids as well as most MCOs, the secularly forced eccentricity vector is directed approximately along the perihelion direction of Jupiter, $\varpi_{\jupiter}=15^\circ$.  Using Laplace--Lagrange secular theory~\citep{Murray:1999}, we calculate that the forced eccentricity vector is directed towards $6^\circ$ ecliptic longitude for semimajor axis $a=3$~AU; it gradually decreases to $344^\circ$ for $a = 2.3$~AU and to $305^\circ$ for $a = 2$~AU where the forced eccentricity vector becomes discontinuous (in the vicinity of the $\nu_6$ secular resonance).  Because the apsidal precession rate is slowest when an asteroid's free eccentricity vector is aligned with the forced eccentricity vector, the non-uniform precession rate over secular timescales leads to the non-uniform distribution of perihelion longitude.  Similar non-uniformities are found for the Amors \cite[see their Figures 14--16]{JeongAhn:2014}.   

The peak direction and the degree of concentration of the $\varpi$ distribution are both different for the high and low inclination MCOs (Fig.~\ref{LNP2}).  The low inclination MCOs ($i<15^\circ$) have a narrow peak centered at $\bar{\varpi} = 341^\circ$ while the high inclination MCOs ($i>15^\circ$) show a broader peak centered at $\bar{\varpi} = 48^\circ$.   The non-uniformity of $\varpi$ distribution of the MCOs plays an important role in the seasonal variation of the impact frequency on Mars.  

%%%%%%%%%%%%%%%%%%%%%%%%%%%%%%%%%%%%%%%%%%%%%%%%%%%%%%%%%%%%%%%%%%%%%%%%%
\begin{figure}[h]
\centering
  \includegraphics[width=300px]{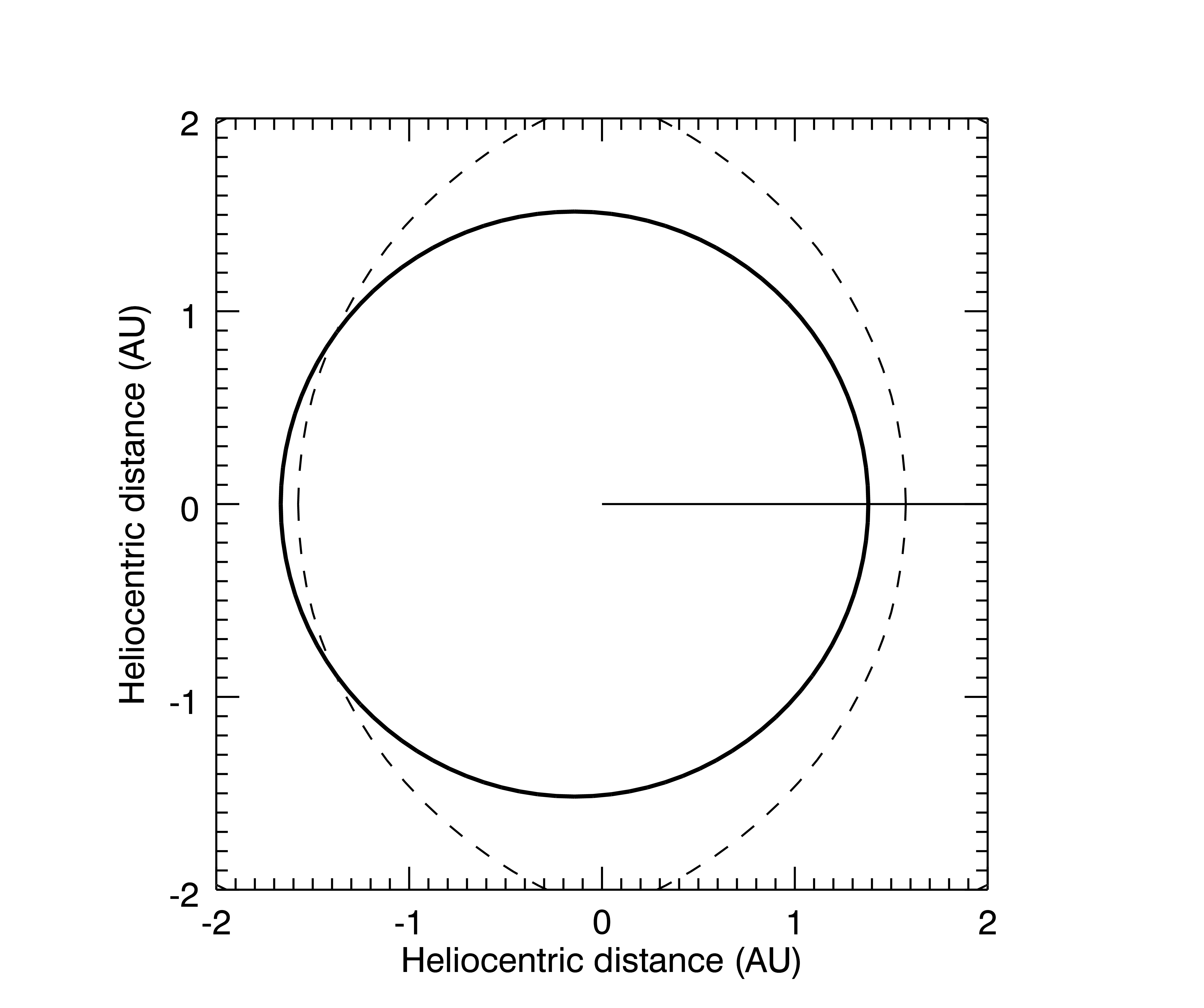}
  \caption{\sm Schematic diagram of Mars' orbit (black solid ellipse) and two possible asteroid orbits (dashed curves) having the same semimajor axis and eccentricity but having different perihelion longitudes, one aligned with Mars' perihelion and the other aligned with Mars' aphelion. The direction of Mars' perihelion is indicated by the solid horizontal line starting from the location of the Sun at the origin. The semimajor axis and eccentricity of the asteroids were chosen as the median values of the bright MCOs ($H<16$), $a=2.41$~AU and $e =0.345$.  For simplicity, all the orbits are assumed to have zero inclination from the projected plane.}
\label{SchD}
\end{figure}
%%%%%%%%%%%%%%%%%%%%%%%%%%%%%%%%%%%%%%%%%%%%%%%%%%%%%%%%%%%%%%%%%%%%%%%%%

Mars' longitude of perihelion also generally shares the secular behavior of the low inclination MCOs; its apsidal precession rate is slowest when $\varpi_{\mars}\approx\varpi_{J}$ (where it has correspondingly longer residence time).  As noted above in Section~\ref{s:defns}, at the present time, Mars' eccentricity is close to its long term maximum; its apsidal precession rate is close to its long term minimum, and its current longitude of perihelion, $\varpi_{\mars} \simeq 336^\circ$, is also close to the peak values of both MCO populations.  
This near-coincidence has significant implications for the current impact rate on Mars because some orbital geometries of MCOs can prohibit impact on Mars.  We illustrate this in Fig.~\ref{SchD}, in which we trace the elliptical orbit of Mars and the orbits of two representative MCO orbits.  The two MCO orbits have the same semimajor axis and eccentricity, which we chose to be median values of the bright MCO sample, $a=2.41$~AU, $e=0.345$, but one has its perihelion aligned with Mars' ($\varpi\approx\varpi_{\mars}$) while the other has perihelion anti-aligned with Mars' ($\varpi\approx\varpi_{\mars}+180^\circ$).  It is obvious that the former prohibits impact on Mars.   Because Mars' current perihelion direction  coincides with the direction of high concentration of the MCOs' perihelia, the intrinsic impact rate can be significantly lower compared to the rate calculated from an assumed uniform random distribution of $\varpi$.  Previous studies have usually adopted a uniform random distribution of $\varpi$.

%%%%%%%%%%%%%%%%%%%%%%%%%%%%%%%%%%%%%%%%%%%%%%%%%%%%%%%%%%%%%%%%%%%%%%%%%
\begin{figure}[h]
\centering
  \includegraphics[width=300px]{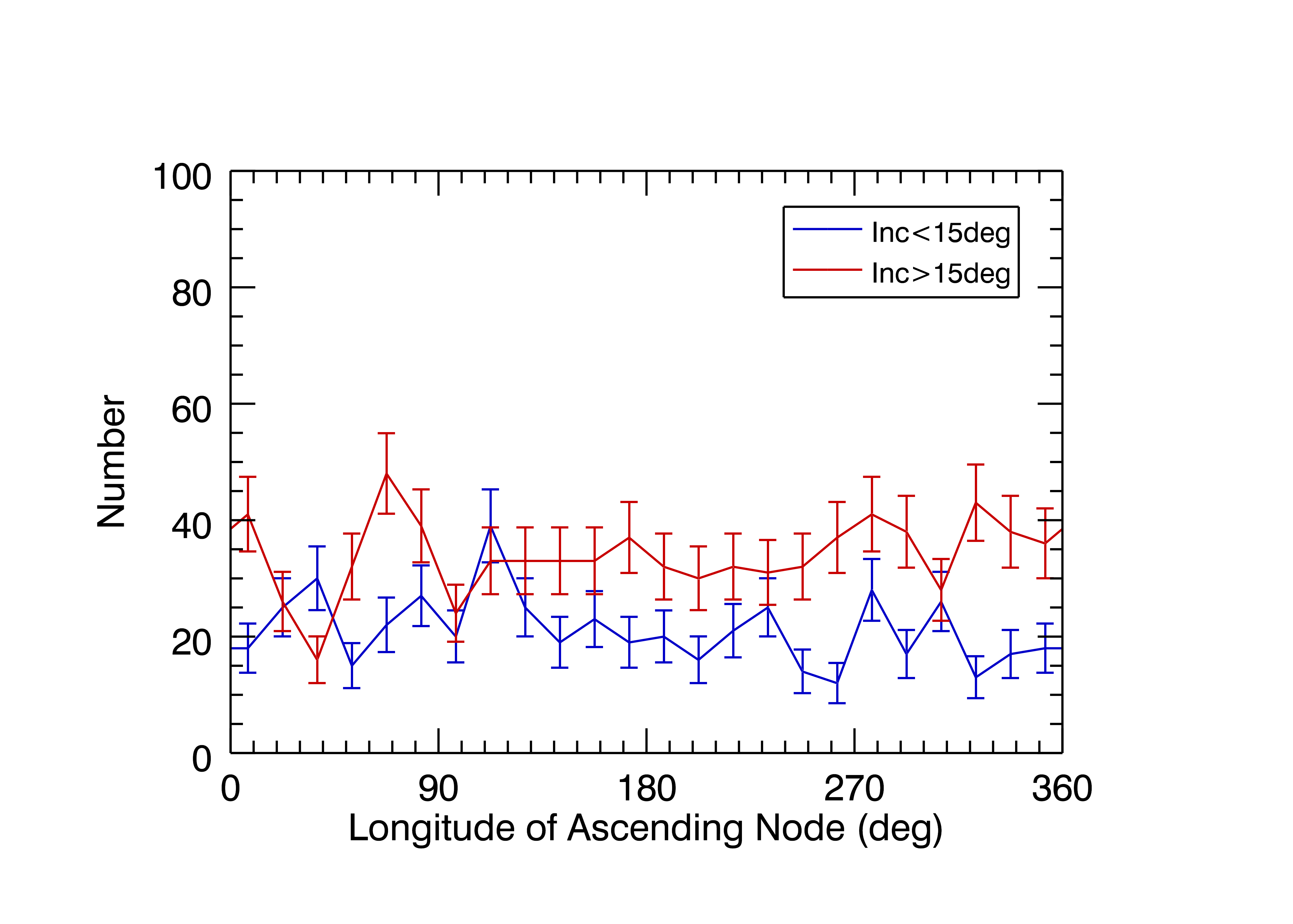}
  \caption{\sm The distribution of the longitude of ascending node, $\Omega$, of bright MCOs ($H<16$), in two different inclination regimes. Both groups are statistically indistinguishable from a uniform distribution. }
\label{NODE}
\end{figure}
%%%%%%%%%%%%%%%%%%%%%%%%%%%%%%%%%%%%%%%%%%%%%%%%%%%%%%%%%%%%%%%%%%%%%%%%%
   
The longitude of ascending node, $\Omega$, of MCOs is uniformly distributed, as shown in Figure~\ref{NODE} for both the low-inclination and the high-inclination populations.  The Rayleigh test for uni-directionality for the bright MCOs indicates that their $\Omega$ distribution is indistinguishable from a uniform distribution ($p>0.05$).  Even though minor intrinsic non-uniformities might be present due to secular dynamics (because the normal to Mars' orbital plane deviates from that of the invariable plane as well as from the local secularly forced inclination vector of MCOs), a minor non-uniformity would not have a significant effect on the impact rate.

\section{Modified {\"O}pik-Wetherill Method}\label{s:opik-wetherill}

We first briefly describe the calculation of the probability per unit time that an MCO in a given orbit approaches Mars within a prescribed encounter distance, then we describe our method for the generation of ``clones'' of MCOs, and finally our method for calculating the impact flux on Mars.

We adopt the method pioneered by~\citet{Opik:1951} and \citet{Wetherill:1967} for the calculation of collision probability, $P$, between two objects on fixed Keplerian orbits.  
While \citet{Opik:1951}'s method restricted one of the colliding pair to a circular orbit, \citet{Wetherill:1967} generalized this method to two eccentric orbits having arbitrary mutual inclination and apsidal orientations. A number of subsequent works have adopted Wetherill's equation for $P$ and provided improved methods for computing its mean value over secular precession timescales~\citep{Greenberg:1982,Bottke:1993,Pokorny:2013}. 

Consider a particular location of Mars in its orbit.  
Following \citet{Wetherill:1967}, the probability per unit time for an MCO to approach Mars within a prescribed distance $\tau$ can be calculated in terms of the relative velocity vector and the orbital periods of the two orbits.  We express Wetherill's formula in simpler notation as
%%%%%%%%%%%%%%%%%%%%%%%%%%%%%%%%%%%%%%%%%%%%%%%%%%%%%%%%%%%%%%%%%%%%%%%%%
\begin{equation}
P(\tau)  = \Bigg\{\begin{array}{lll}
& \pi \tau\big(2  U_0 T T_0 \sqrt{1 -  ({\bf U}\cdot{\bf U_0})^2U^{-2}U_0^{-2}}\big)^{-1} &\hbox{if $\tau>D_{\min}$},\\
& 0& \hbox{if $\tau<D_{\min}$},
\end{array}\label{WethPX}
\end{equation}

%%%%%%%%%%%%%%%%%%%%%%%%%%%%%%%%%%%%%%%%%%%%%%%%%%%%%%%%%%%%%%%%%%%%%%%%%
where $D_{min}$ is the minimum distance between the orbit of the MCO and the orbit of Mars, ${\bf U}_0$ is the heliocentric velocity of Mars at the encounter, ${\bf U}$ is the encounter velocity of the MCO relative to Mars, $T$ is the orbital period of the MCO, and $T_0$ is the orbital period of Mars.  %Angle $\alpha$ is the angle between motion of Mars and the x axis and $\tau$ is the colliding radius. 
This expression for $P(\tau)$ forms the backbone for our impact flux calculations.  The dependence of $P(\tau)$ on the orbital elements of Mars and of the MCO is implicit in the velocities, ${\bf U},{\bf U}_0,$ in the right hand side of Eq.~\ref{WethPX}.   Note that this probability does not account for any gravitational focusing effect of Mars, which we will address below. 

Previous calculations~\citep{Opik:1951,Wetherill:1967,Greenberg:1982,Bottke:1993} commonly adopted uniform  distributions of the angular orbital elements by assuming uniform angular precession rates to calculate impact probabilities. However, this is not justified for MCOs, as shown in the previous section. Furthermore, these distributions differ for the high and low inclination subgroups of MCOs.  Therefore, we carry out  impact flux calculations separately for each subgroup.  

In order to avoid the small number statistics problem, we generate clones having the current orbital distribution of MCOs, as follows.   
For each MCO within each inclination subgroup, we generate ten thousand clones by taking each triad, $(a,e,i)$ and assigning to it a value of $\omega$ and of $\varpi$ chosen randomly from the $\omega$'s and $\varpi$'s of the low inclination (respectively, high inclination) subgroup of MCOs.   In this way, we preserve the correlations amongst $(a,e,i)$ as well as the respective non-random distributions of $\omega$ and of $\varpi$ of the high and low inclination subgroups, while generating a sufficient number of clones to obtain good statistics for the impact flux calculations.  We understand that, even though we subdivided the MCOs into two distinct inclination groups, there could be residual correlations between $\omega$, $\varpi$ and the other orbital elements; we neglect this complication here.

For each MCO clone, we calculate the Minimum Orbit Intersection Distances (MOIDs) between Mars and the clone, with the code developed by  \cite{Gronchi:2005}.  This distance is defined as the closest possible distance between any random locations on each orbit. The MOID is a widely used measure for assessing the impact hazard of near-Earth objects on Earth but can be used with any two osculating orbits.  We identify a possible collision between two orbits if their MOID is less than a prescribed small distance, $D$.  We use $D=0.001$~AU, which corresponds to 44 times the physical radius of Mars, $R_{\mars}$. 
(A correction factor, $(R_{\mars}/D)^2$, is applied later (see below) to convert to actual collisional statistics.)  We record the velocity vectors of the MCO and of Mars at the MOID location, to enable the calculation of collision probability, $P(D)$, with Eq.~\ref{WethPX}.  We also record the longitude where MOID occurs, to track the seasonal variation of the impact rate.   It is theoretically possible that two orbits have two ``crossings'' at two nodes. However, such cases are very rare, and we neglect these rare possibilities, recording only a single position where MOID occurs. 

Now, among all the MCO clones, only a small fraction meets the criterion of MOID$<D$. These MCOs would potentially approach Mars closer than the small distance $D$ %0.001~AU 
in the near future (absent any orbit changes, e.g. owing to planetary perturbations).  
For each of these potential Mars colliders, we calculate the probability $P(D)$ with the use of Eq.~\ref{WethPX}.  Then, the impact rate of the bright MCOs of absolute magnitude $H<H_{lim}$ is computed as follows: 
%%%%%%%%%%%%%%%%%%%%%%%%%%%%%%%%%%%%%%%%%%%%%%%%%%%%%%%%%%%%%%%%%%%%%%%%%
\begin{equation}
I(H_{lim}) = \frac{ (1+b)R_{\mars}^2}{D^{2} N_c} {\sum_i \left( 1+ \frac{2GM_{\mars}}{R_{\mars} U_i^2} \right) P_i (D)},
\label{impactrate}\end{equation}
%%%%%%%%%%%%%%%%%%%%%%%%%%%%%%%%%%%%%%%%%%%%%%%%%%%%%%%%%%%%%%%%%%%%%%%%%
where the sum is over all the cloned population of MCOs and $N_c=10^4$ is the clone multiplicity.  There are three correction factors included above: $(1+b)$ accounts for the observational incompleteness of our sample of bright MCOs (in Section~\ref{s:Hdist}, we estimated $b=0.35$ for high inclination MCOs and $b=0.11$ for the low inclination ones, respectively, for $H_{lim}=16$); the factor $(R_{\mars}/D)^2$ accounts for the inflated collision cross section in our calculation of $P_i(D)$; the factor $(1+2GM_{\mars}(R_{\mars}^{-1} U_i^{-2})$ accounts for the gravitational enhancement of the collision cross section of Mars for an MCO having encounter velocity $U_i$. 

The impact flux on Mars is defined as the number of impacts per unit time per unit surface area, i.e. $F = I/(4\pi R^2_{\mars})$, therefore,
\begin{equation}
F(H_{lim}) = \frac{(1+b)}{4\pi D^{2} N_c} {\sum_i \left( 1+ \frac{2GM_{\mars}}{R_{\mars} U_i^2} \right) P_i (D)}.
\label{impactflux}\end{equation}

Because we also recorded the longitude where the MOID occurs for each cloned MCO, we can calculate the impact flux on Mars as a function of its longitude (measured from Mars' perihelion direction), as follows.  Consider the longitude bin $\{\lambda_j - \frac{1}{2}\Delta \lambda, \lambda_j+\frac{1}{2}\Delta\lambda\}$. The impact rate in this bin is given by

%%%%%%%%%%%%%%%%%%%%%%%%%%%%%%%%%%%%%%%%%%%%%%%%%%%%%%%%%%%%%%%%%%%%%%%%%
\begin{equation}
F_{j} (H_{lim}) =  \frac{(1+b)T_0}{4\pi D^{2} N_c\Delta T_j } {\sum_{i} \left( 1+ \frac{2GM_{\mars}}{R_{\mars} U_i^2} \right)  P_i (D)} \delta_{ij},
\label{Plmb2}
\end{equation}
%%%%%%%%%%%%%%%%%%%%%%%%%%%%%%%%%%%%%%%%%%%%%%%%%%%%%%%%%%%%%%%%%%%%%%%%%
where $\delta_{ij}$ is unity when the impact site of the $i$-th clone is located within the $j$-th longitude bin, and zero otherwise, and $\Delta T_j/T_0$ is the fraction of its orbital period that Mars spends in the $j$-th longitude bin.  The following relation holds by definition,
%%%%%%%%%%%%%%%%%%%%%%%%%%%%%%%%%%%%%%%%%%%%%%%%%%%%%%%%%%%%%%%%%%%%%%%%%
\begin{equation}
F(H_{lim})  = {\sum_{j} F_{j} (H_{lim}) \frac{\Delta T_j}{T_0}}.
\label{QH}
\end{equation}
%%%%%%%%%%%%%%%%%%%%%%%%%%%%%%%%%%%%%%%%%%%%%%%%%%%%%%%%%%%%%%%%%%%%%%%%%

In order to assess the statistical uncertainty in our impact flux estimates, we carry out the impact flux calculations with three independent realizations of the randomly generated clone sets.  In the results reported below, we quote the average over these three sets.  The standard deviation of these three realizations is our estimate of the statistical uncertainty due to the random generation of clones; we find this to be on the order of a efew percent of the average.  Additionally, in order to illustrate the role of the non-uniformity of angular distributions, we also generate a ``control set'' of clones by assigning uniform random values to $\omega$ and $\varpi$, and we carry out the impact flux calculation separately with this set.

\section{Results}\label{s:results}

\subsection{Impact flux on Mars}\label{s:impactflux}
%%%%%%%%%%%%%%%%%%%%%%%%%%%%%%%%%%%%%%%%%%%%%%%%%%%%%%%%%%%%%%%%%%%%%%%%%
\begin{table}
  \caption{\sm The impact flux of bright MCOs ($H<16$) on Mars and its seasonal variation.  }
\begin{tabular}{|l||c|c|c|c|c|}
\hline
 & \multicolumn{3}{|c|}{mean impact flux, $F(16)$} &\multicolumn{2}{|c|}{}\\
$\omega,\varpi$ distribution & \multicolumn{3}{|c|}{$(10^{-23} \rm{km}^{-2} \rm{s}^{-1}$)}  &\multicolumn{2}{|c|}{aphelion-to-perihelion flux ratio} \\
 \cline{2-6}
& all &  $i>15^\circ$ & $i<15^\circ$ & $\pm30^\circ$  & half Martian year  \\
  \hline  \hline
  Non-uniform  & 2.9 & 1.4 & 1.6  & 4.3  & 2.1  \\
  Uniform  & 6.1 & 2.4 & 3.8 & 15.4 & 5.4  \\
  \hline
   &\multicolumn{5}{|l|}{Gravitational focusing neglected} \\
  \hline
  Non-uniform & 2.1 & 1.2 & 0.9 & 3.8 & 1.9  \\
  Uniform  & 4.1 & 2.0 & 2.1 & 11.5 & 4.4  \\
\hline 
\end{tabular}
\label{TBrate}
\end{table}
%%%%%%%%%%%%%%%%%%%%%%%%%%%%%%%%%%%%%%%%%%%%%%%%%%%%%%%%%%%%%%%%%%%%%%%%%

The impact flux and its seasonal variation are calculated following the steps explained in the previous section.  As noted, we carried out calculations for the high inclination and the low inclination groups of MCOs separately, and then combined the two sets.  The detailed results are listed in Table~\ref{TBrate}.   For comparison, the results from the case of uniform angular distribution are also listed in Table~\ref{TBrate}.  In addition,  we intentionally neglected the gravitational focusing in order to illustrate the role of gravitational focusing in Martian impact statistics; we list the results so obtained in the last two rows in the Table.   All of these listed results, we report the average of three simulations; the standard deviations of the fluxes obtained in these simulations are all below 3.5\% of the reported flux. 
 We describe first our results for the mean impact flux~($F$, Eq.~\ref{impactflux}) and then our results for the seasonal variation of the impact flux ($F_j$, Eq.~\ref{Plmb2}).  The distribution of impact velocity and the effect of gravitational focusing will be described afterwards.

We find that the overall mean impact flux, $F(16)$~(Eq.~\ref{impactflux}), is $2.9 \times 10^{-23}\rm{km}^{-2} \rm{s}^{-1}$.  The contributions to this mean flux by the two inclination subgroups are similar: $1.4  \times 10^{-23}\rm{km}^{-2} \rm{s}^{-1}$ and $1.6 \times 10^{-23}\rm{km}^{-2} \rm{s}^{-1}$ for the high inclination MCOs and the low inclination groups, respectively.   Although the population of the high inclination subgroup is twice as large as the  low inclination subgroup,  the low intrinsic impact probabilities of higher inclination MCOs reduces their contribution to the impact statistics. 
It is noteworthy that the impact rates are significantly higher when we adopt a uniform random distribution of the angular elements ($\omega,\varpi$): the impact flux for the high inclination group is larger by a factor 1.7 and the flux for the low inclination group is larger by a factor 2.4; the overall rate is 2.1 times higher than for the non-uniform case. 
The difference is larger for the low inclination group.  We can understand these trends with the help of the schematic diagram in Figure~\ref{SchD} which illustrates that MCOs whose eccentricity vectors are aligned with Mars' eccentricity vector are less likely to impact Mars.  For the high inclination group, the axial distribution of $\omega$ also plays an important role in reducing the impact flux because these MCOs tend to reach perihelion when they are located off the orbital plane of Mars. 

%%%%%%%%%%%%%%%%%%%%%%%%%%%%%%%%%%%%%%%%%%%%%%%%%%%%%%%%%%%%%%%%%%%%%%%%%
\begin{figure}[h]
\centering
  \includegraphics[width=300px]{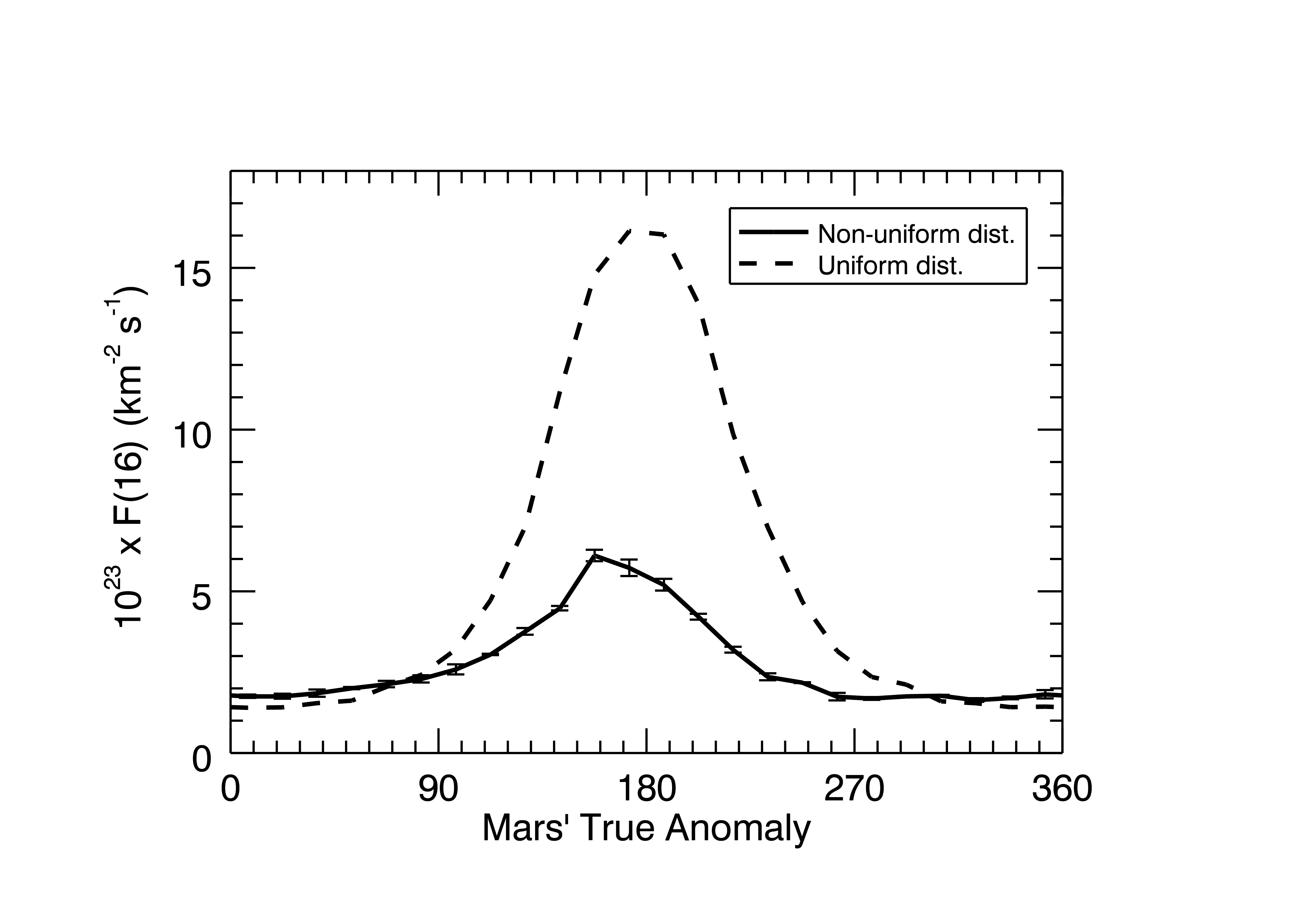}
  \caption{\sm (Solid line) The seasonal variation of the impact flux, $F_j(16)$ (Eq.~\ref{Plmb2}), of the bright MCOs $(H<16)$; for each longitude bin, we calculated the mean from three independent realizations of the randomly generated clones of MCOs; the error bars indicate the standard deviation of the three simulations.  (Dashed line) For comparison, we also computed the impact flux with an assumed uniform distribution of the angular elements ($\omega,\varpi)$ of MCOs.  The longitude bin size is 15 degrees.
}
\label{Season}
\end{figure}
%%%%%%%%%%%%%%%%%%%%%%%%%%%%%%%%%%%%%%%%%%%%%%%%%%%%%%%%%%%%%%%%%%%%%%%%%

%%%%%%%%%%%%%%%%%%%%%%%%%%%%%%%%%%%%%%%%%%%%%%%%%%%%%%%%%%%%%%%%%%%%%%%%%
\begin{figure}[h]
\centering
  \includegraphics[width=300px]{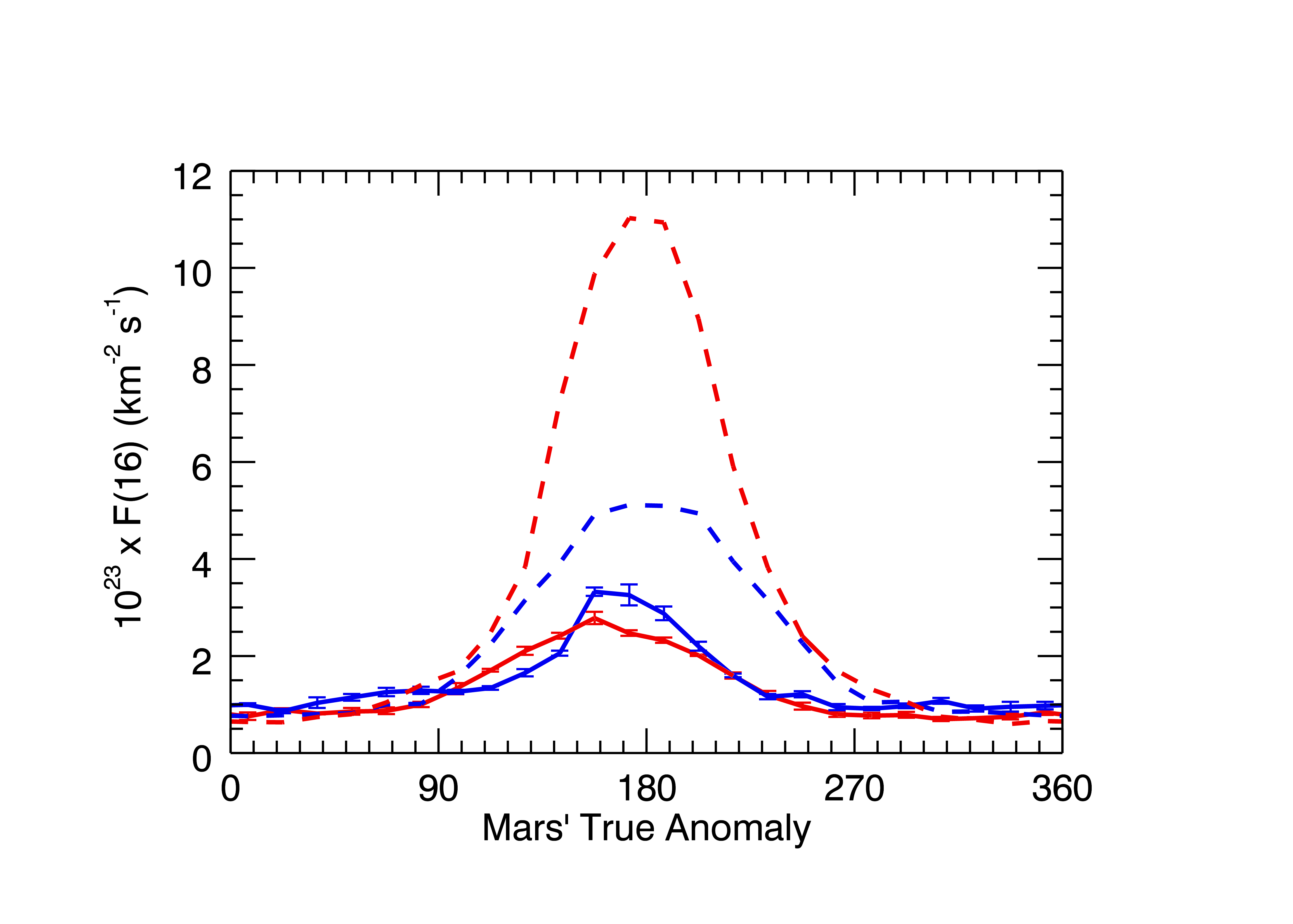}
  \caption{\sm Same as Figure~\ref{Season}, but limited to the high inclination subgroup, $i>15^\circ$  (red), and the low inclination subgroup, $i<15^\circ$ (blue).
}
\label{SeasonHL}
\end{figure}
%%%%%%%%%%%%%%%%%%%%%%%%%%%%%%%%%%%%%%%%%%%%%%%%%%%%%%%%%%%%%%%%%%%%%%%%%

The results for the seasonal variation in the impact flux are shown in Figure~\ref{Season} and~\ref{SeasonHL}, in which we plot the total impact flux, and the impact flux contributed by the low inclination MCOs and the high inclination MCOs, respectively, as a function of Mars' true anomaly, i.e., $F_j$~(Eq.~\ref{Plmb2}), with 15-degree longitude bins. 
The continuous line curves plot the results for the case of the non-uniform distribution of the angular elements ($\omega,\varpi$) of the bright MCOs.  For comparison, we also plot the result for an assumed uniform random distribution of these angles (dashed line).  We observe that the impact flux on Mars peaks when the planet is near aphelion and MCOs are at their perihelion.  There are at least two reasons for this: the orbit-crossing geometry (see Figure~\ref{SchD}) is more favorable, and the spatial density of MCOs increases with heliocentric distance.  A third reason is due to the lower encounter velocity (hence larger gravitational cross section) of MCOs; this is described below in Section~\ref{s:impactvelocities}.

It is also evident that the non-uniform distribution of the angular elements significantly affects the amplitude of the seasonal variability of the impact flux.  To express this quantitatively, we integrated the impact flux over the true anomaly range of $\pm 30^\circ$ centered at Mars' aphelion and and its perihelion.   Then we find that the aphelion-to-perihelion ratio of the impact flux is 15.4 for the case of the uniform distribution but it is only 4.3 for the actual non-uniform distribution.  Another way to illustrate this is with the half-year ratio: integrating over half a Martian year centered at the planet's aphelion, we find that impacts are 2.1 times more frequent than during the half Martian year centered at perihelion.  In contrast, this ratio is 5.4 if we assume uniform random distribution of the angular elements.  The non-uniform distribution of the MCOs' longitude of perihelion, with a concentration near Mars perihelion, means that the orbital geometry that favors impact conditions (Figure~\ref{SchD}) occurs less frequently (compared to the case of a uniform random distribution of $\varpi$), thereby diminishing the seasonal concentration of the impacts near Mars' aphelion and also diminishing the overall impact flux.

We remark that the high and low inclination subgroups have a similar seasonal variation of the impact flux  when we take account of the non-uniform angular distribution.  However, we find a significant difference in their seasonal variation in the simulations with uniform angular distributions:  the ($\pm30^\circ$) aphelion-to-perihelion ratio is about 9 for the high inclination subgroup and about 25 for the low inclination subgroup.  This may be of importance for the impact flux of small MCOs which may be dominated by low inclinations and possibly more uniform angular distribution due to non-gravitational effects.  We discuss this further in Section~\ref{s:smallimpactors}.

%%%%%%%%%%%%%%%%%%%%%%%%%%%%%%%%%%%%%%%%%%%%%%%%%%%%%%%%%%%%%%%%%%%%%%%%%
\begin{figure}[h]
\centering
\includegraphics[width=300px]{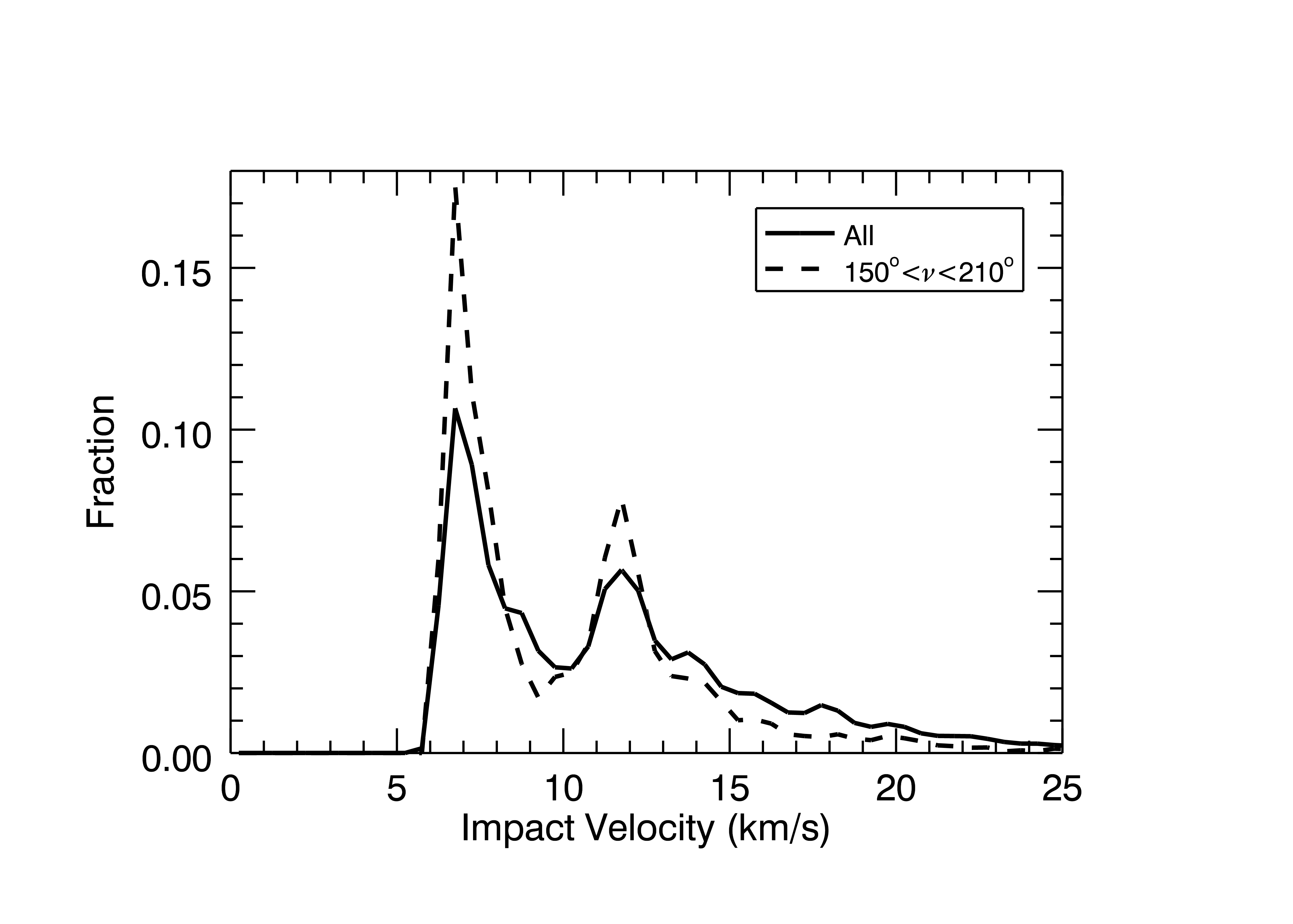}
  \caption{\sm The solid line plots the distribution of impact velocity of bright MCOs with Mars, the  dashed line plots the impact velocity distribution near Mars' aphelion (true anomaly in the range $150^\circ$ to $210^\circ$).  We accounted for the effects of the intrinsic collision probability of individual MCOs as well as the different observational completeness levels between the high inclination population and low inclination population. }
\label{ImpVel}
\end{figure}
%%%%%%%%%%%%%%%%%%%%%%%%%%%%%%%%%%%%%%%%%%%%%%%%%%%%%%%%%%%%%%%%%%%%%%%%%

\subsection{Impact velocities}\label{s:impactvelocities}

As mentioned previously, in our simulations we tracked the encounter velocities, $v_{enc}$, of MCOs with Mars.   We computed the impact velocities as $v_{imp}^2= v_{enc}^2+2GM_{\mars}R_{\mars}^{-1}$. In Figure~\ref{ImpVel}, we plot the distribution of impact velocity of MCOs with Mars integrated over the Martian year (solid line curve), and the  impact velocity distribution when the planet is near aphelion (dashed curve).  For the latter, we integrated over the true anomaly range of $\pm 30^\circ$ around the aphelion of Mars. 

The  mean encounter velocity is 10.1 km/s, equivalent to a mean impact velocity of 11.5 km/s.  (The simulations with uniform random distribution of the angular elements give slightly lower values,  8.5 km/s and 10.1 km/s, respectively.)  The impact velocity distribution has two distinct peaks, at $\sim6.5$~km/s and $\sim11.5$~km/s. The lower velocity peak is contributed by low inclination MCOs, and the higher velocity peak is by high inclination ones, because of their larger vertical component of the relative velocity. This justifies our choice in separating the two groups with the inclination boundary at $i=15^\circ$.  (These two peaks can be also found in the results of \citet{LeFeuvre:2011}, in their Fig.~4.)  The overall encounter velocity distribution is similar to the distribution from aphelion colliders (shown as the dashed curve in Figure~\ref{ImpVel}) because the aphelion collisions dominate the impact flux. 

\begin{figure}[h]
\centering
  \includegraphics[width=300px]{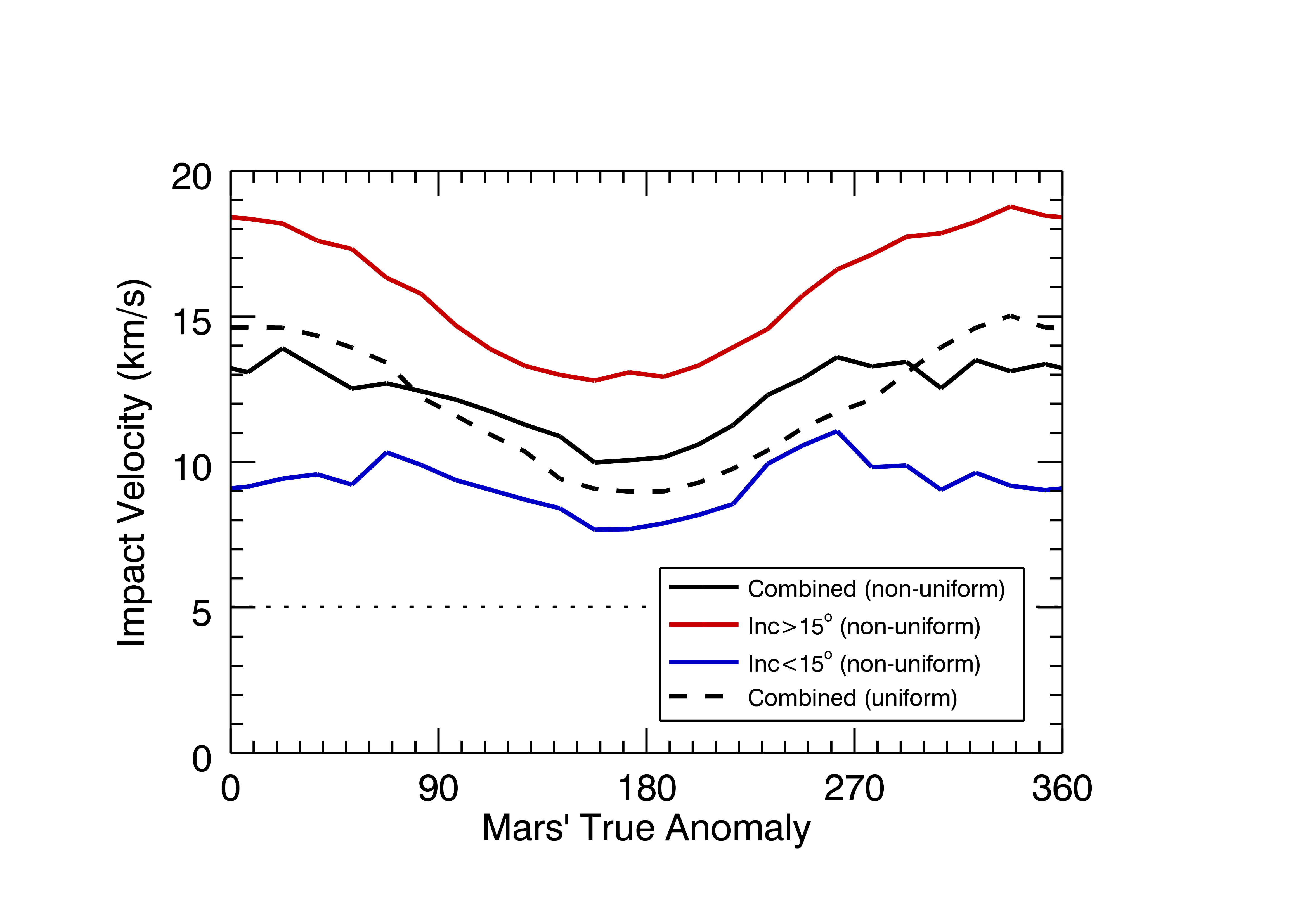}
  \caption{\sm Mean impact velocity of bright MCOs ($H<16$) and its seasonal variation.  The red and blue curves plot the mean impact velocity of the high inclination subgroup ($i>15^\circ$) and the low inclination subgroup ($i<15^\circ$), respectively, and the continuous black curve plots the combined mean encounter velocity of all bright MCOs.  For comparison, the dashed black curve shows the mean impact velocity resulting from the uniform distribution of angular elements. The dotted horizontal line indicates the escape velocity of Mars, $v_{esc} = 5~\rm{km/s}$.
  }
  \label{ImpVLng}
\end{figure}

The impact velocities are seasonally variable.  In Figure~\ref{ImpVLng}, we plot the mean impact velocity of MCOs with Mars, as a function of Mars' true anomaly (with 15--degree longitude bins).   The mean impact velocity of the high and low inclination subgroups of MCOs as well as that of the combined population are plotted.   The mean impact velocity is lowest near Mars' aphelion and highest when the planet is near perihelion.  The seasonal variation is larger for the high inclination group.  Our calculations also show that the seasonal variation for the case of non-uniform angular distribution (black solid line) is subdued compared to the case of uniform random angular distribution (black dashed line). 
 
 The encounter velocity affects the gravitational focusing factor in the impact flux calculation. Table~\ref{TBrate} shows that gravitational focusing not only increases the mean impact flux by 40\% but also its seasonal variability. The low mean encounter velocity near Mars' aphelion enhances the collision cross section more than near perihelion, and this is an additional factor that enhances the impact flux near Mars' aphelion.
Also it is evident that the effect of gravitational focusing is much more significant for low inclination MCOs due to their lower encounter velocities, accounting for about 70\% increased impact flux.

%%%%%%%%%%%%%%%%%%%%%%%%%%%%%%%%%%%%%%%%%%%%%%%%%%%%%%%%%%%%%%%%%%%%%%%%%

%%%%%%%%%%%%%%%%%%%%%%%%%%%%%%%%%%%%%%%%%%%%%%%%%%%%%%%%%%%%%%%%%%%%%%%%%
%%%%%%%%%%%%%%%%%%%%%%%%%%%%%%%%%%%%%%%%%%%%%%%%%%%%%%%%%%%%%%%%%%%%%%%%%
\begin{figure}[h]
\centering
\includegraphics[width=300px]{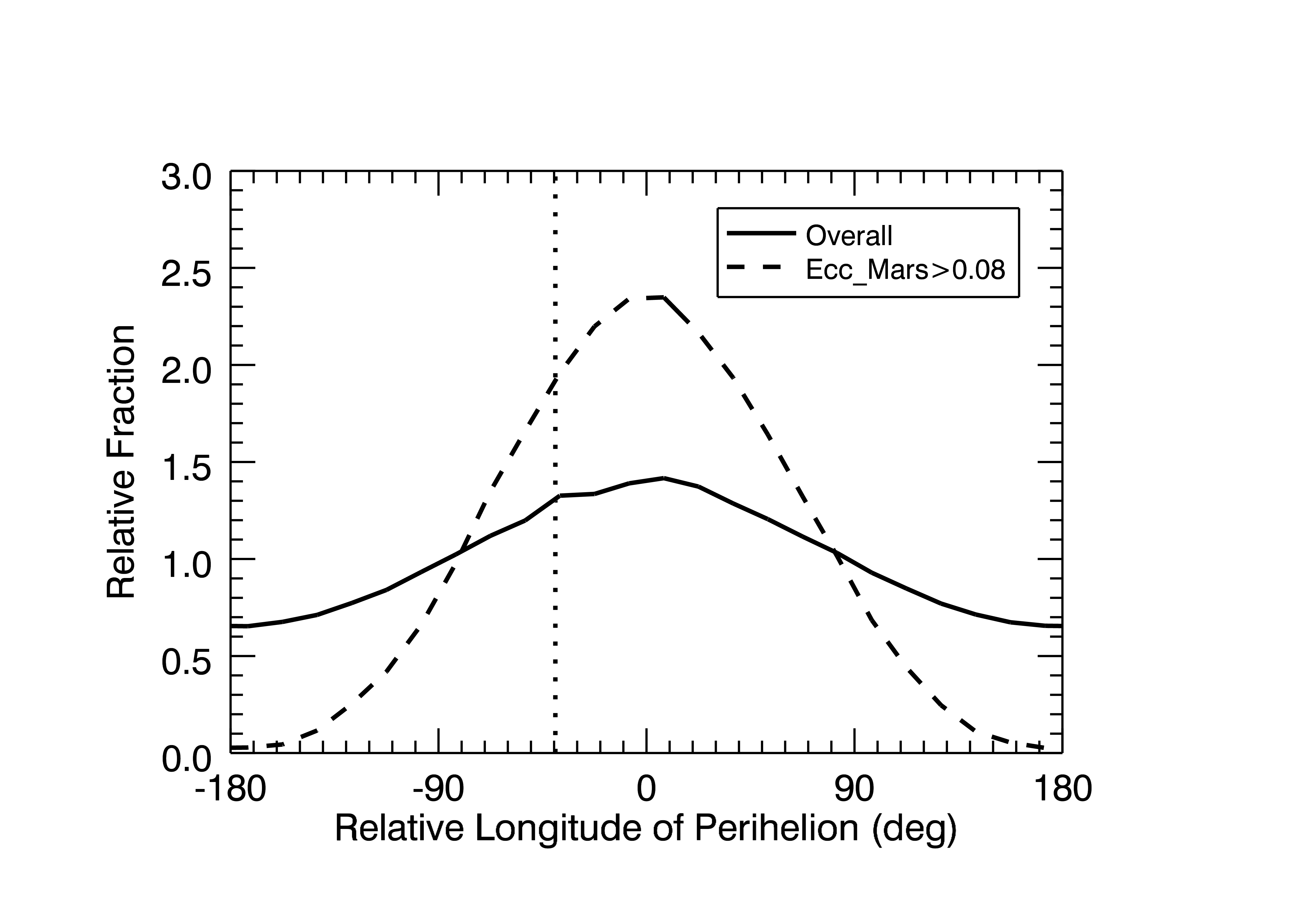}
  \caption{\sm The time-weighted distribution of the longitude of Mars' perihelion relative to Jupiter's, $\varpi_{\mars}-\varpi_{\jupiter}$, in two different regimes of Mars' eccentricity. The calculation was done with an orbital integration of all eight planets for $10^8$ years. The vertical dotted line indicates the current value of $\varpi_{\mars}-\varpi_{\jupiter}$.}
  \label{RelLNP}
\end{figure}
%%%%%%%%%%%%%%%%%%%%%%%%%%%%%%%%%%%%%%%%%%%%%%%%%%%%%%%%%%%%%%%%%%%%%%%%%

%%%%%%%%%%%%%%%%%%%%%%%%%%%%%%%%%%%%%%%%%%%%%%%%%%%%%%%%%%%%%%%%%%%%%%%%%
\subsection{Observational sample bias}\label{s:samplebias}

The results described above are based on the orbital distribution of the bright ($H<16$) MCOs, together with an estimate of their completeness.  We repeated the calculations, for the MCOs with a brighter magnitude cut-off, $H<15$, for which the known population is more complete. This allows us to examine whether the observational bias in the orbital distribution of fainter objects has an effect on our results.  There are only 516 MCOs having $H<15$.  With this smaller set, the detailed results varied from the results from $H<16$ objects, but we still find the same ratio, 2.1, of the impact flux in the half-Martian year centered at the planet's aphelion versus perihelion. The mean impact velocity, 11.6 km/s, also matches the result from $H<16$ objects.   We find their mean impact flux is $F(15)=7.6\times 10^{-24} \rm{km}^{-2} \rm{s}^{-1}$.   Extrapolating to the fainter population of $H<16$ MCOs by use of Eq.~\ref{cumdist}, we then estimate $F(16)=2.5\pm0.4\times 10^{-23} \rm{km}^{-2} \rm{s}^{-1}$.  This is within about 15\% of our simulation result (see Table~\ref{TBrate}), and it indicates the uncertainty owed to the observational sample bias.

%%%%%%%%%%%%%%%%%%%%%%%%%%%%%%%%%%%%%%%%%%%%%%%%%%%%%%%%%%%%%%%%%%%%%%%%%

\subsection{Mars/Moon and Mars/Earth impact flux ratios}\label{s:ratio}

Studies of crater chronology based on impact crater densities often use estimates of the impact flux ratio of Mars and the Moon.   Additionally, in the next section we will make use of the impact flux ratio of Mars and the Earth.  We therefore applied the same method as above to calculate the impact flux of bright Earth-crossing objects (ECOs) on the Moon and Earth. 

Making use of the same source of minor planet data (the MPCORB dataset), we identified the 516 bright ECOs having $H<18$; these are assumed to be observationally complete \citep{Mainzer:2011,JeongAhn:2014}. Most of this population of ECOs is the Apollo subgroup of NEOs.  The Apollos have a strongly axial distribution of the argument of perihelion, concentrated near 0 and 180 degrees, but their longitude of ascending node is statistically indistinguishable from a uniform random distribution~\citep{JeongAhn:2014}.  Therefore, we took account of the non-uniformity in $\omega$ and  carried out the calculation of the impact flux for the Earth and the Moon in the same way as we did for the Martian impact flux.  We included gravitational focusing for individual encounters, but for the lunar impact flux we neglected the orbital velocity of the Moon about the Earth.  Our result for the mean impact flux on the Moon is $F_{\!\!\rightmoon}(18)=6.06 \times 10^{-23}\rm{km}^{-2} \rm{s}^{-1}$ and for the Earth it is $F_{\oplus}(18)=1.06\times 10^{-22}\rm{km}^{-2} \rm{s}^{-1}$; the Earth/Moon impact flux ratio is $1.7$.

To obtain the Mars/Moon and Mars/Earth impact flux ratios, we extrapolate our Martian impact flux estimate, $F(16)$, to the fainter magnitude, $H<18$, by using a multiplication factor, $N_{H<18}/N_{H<16}$. We calculate $N_{H<18}$ and $N_{H<16}$ from Eq.~\ref{cumdist}, and obtain $F(18) = 3.0 \times 10^{-22} \rm{km}^{-2}\rm{s}^{-1}$ for the Martian impact flux.  From this, we find that the current Mars/Moon impact flux ratio is 5.0, and the Mars/Earth impact flux ratio is 2.8.

%%%%%%%%%%%%%%%%%%%%%%%%%%%%%%%%%%%%%%%%%%%%%%%%%%%%%%%%%%%%%%%%%%%%%%%%%

\subsection{Flux of small impactors}\label{s:smallimpactors}

The flux of small impactors, in the size range 1--10 meters, is of interest because these are likely to be frequent enough that space telescopes currently orbiting Mars would be able to observe them as fresh craters on the surface of the planet.  The orbital distribution of small MCOs is not certain, as they may or may not track the orbital distribution of the bright MCOs.  (Their long term orbital evolution is likely to be  affected by the Yarkovsky effect. However, the mean semimajor axis drift rate for small MCOs of diameter $D\lesssim30$~m is quite small, $|\dot a|\lesssim 0.01$~AU per million years \citep{O'Brien:2005}.) 
If the orbital distribution of small MCOs is the same as that of the bright MCOs, then we can straightforwardly predict that the seasonal variation of their current impact flux will be similar, i.e., a maximum impact flux at aphelion, and a ratio of about 4 for the flux at $\pm30^\circ$ aphelion-to-perihelion.  Should observations find a significantly different (larger) seasonal variation of fresh craters on Mars, it would point to differences in the orbital distribution of the small MCOs relative to the large MCOs. For example, a uniform distribution of the angular parameters of small impactors would produce a much larger aphelion-to-perihelion flux ratio, as Table~\ref{TBrate} suggests.

Here we extend our impact flux calculation to meter-size MCOs, using two separate estimates of the population of small MCOs.  In the first, we extrapolate the $H$--magnitude distribution of the bright MCOs (Eq.~\ref{cumdist}) down to $H$ magnitude of 33 which corresponds to approximately meter-size objects (for albedo of 0.1).  We note that our simple description of the cumulative $H$-magnitude distribution as a combination of two power law functions, with different slopes for the low and high inclination subgroups, suggests that, at the faint magnitude of $H\approx33$, the low inclination objects are about 10 times more numerous than high inclination objects.  Admittedly, this is a very large extrapolation; a small change in the power law slope, e.g., by $\sim0.1$ near $H$ magnitude of 17, can lead to a factor of $\sim20$ over- or under-estimate of the cumulative population of $H<33$ MCOs.  Nevertheless, we bravely proceed, and we find that the low and high inclination subgroups imply a flux of $\sim1.78\times10^{-14}  \rm{km}^{-2} \rm{s}^{-1}$ and $\sim1.26\times10^{-15}  \rm{km}^{-2} \rm{s}^{-1}$, respectively, for an estimated total bolide flux on Mars of $F(33)\approx1.9\times10^{-14}  \rm{km}^{-2} \rm{s}^{-1}$.   

For an alternative and independent estimate, we look to the observational data for the bolide flux at Earth and use our estimate of the Mars/Earth flux ratio.  \citet{Brown:2002} have reported the cumulative size distribution of terrestrial bolides as
\begin{equation}
\log_{10} N (>D) = 1.568 - 2.70 \log_{10} D,
\label{bol}
\end{equation}
where $D$ is the bolide diameter in meters, and $N(>D)$ is the average number of bolides per year.  This indicates about 37 impacts per year on Earth for $D>1$~m (equivalently, $H\lesssim 33$) objects, or $F_\oplus(33)\approx2.3\times10^{-15}  \rm{km}^{-2} \rm{s}^{-1}$ in our notation.  This estimate is consistent with more recent bolide data~\citep{Brown:2013}.  Then, making use of the Mars/Earth impact flux ratio of 2.8 that we calculated in the previous section, and assuming that this ratio is preserved down to meter size objects, the Martian bolide flux is estimated to be $F(33)\approx6.4\times10^{-15}  \rm{km}^{-2} \rm{s}^{-1}$.

The above two estimates imply that there are 103--305 impacts of meter-size objects per year on Mars.  
An impact rate of this magnitude should be readily detectable in the rate of fresh craters.   Moreover, if the distribution of the angular elements of the MCO population is uniform, the seasonal variation of the impact rate would be readily distinguishable from the seasonal variation produced by the non-uniform angular distribution observed for the bright MCOs.   If, as we expect from Eq.~\ref{cumdist}, the meter--size impactors are dominated by low inclination orbits, then we predict that their ($\pm30^\circ$) aphelion-to-perihelion flux ratio is $\sim4.4$ in the case of non-uniform angular distribution, but as high as $\sim25$ in the case of uniform angular distribution.  Therefore, continuing efforts towards detecting fresh craters on the Martian surface will provide not only a better estimate of the mean impact flux but also reveal the orbital distribution of meter size objects, particularly their non-uniformity in angular elements. 
 
Neglecting atmospheric effects, impacts on Mars by meter-size projectiles would produce craters of diameter $\sim 30$~m \citep{Popova:2007}.  \cite{Daubar:2013} report the current observed Martian cratering rate for craters larger than 30 m as $5\times10^{-8} \rm{km}^{-2} \rm{yr}^{-1}$, which is equivalent to about 14 impacts per year. 
Thus, our estimate is about 7 to 21 times higher than this observed value. However, small projectiles in this size range experience non-negligible mass loss and deceleration in velocity while traversing the Martian atmosphere \citep{Williams:2014}, reducing the crater production rate below our estimate of the meter-size impact flux.  

%%%%%%%%%%%%%%%%%%%%%%%%%%%%%%%%%%%%%%%%%%%%%%%%%%%%%%%%%%%%%%%%%%%%%%%%%

\section{Discussion}\label{s:discussion}

%%%%%%%%%%%%%%%%%%%%%%%%%%%%%%%%%%%%%%%%%%%%%%%%%%%%%%%%%%%%%%%%%%%%%%%%%

\subsection{Comparisons with previous studies}\label{s:comparison}
%%%%%%%%%%%%%%%%%%%%%%%%%%%%%%%%%%%%%%%%%%%%%%%%%%%%%%%%%%%%%%%%%%%%%%%%%

We now compare our results to those reported in previous studies.  
Previous studies have not reported detailed calculations of the seasonal variation, therefore we confine our comparisons to the mean impact flux and the impact velocities
 There are two main studies for this comparison, \citet{Ivanov:2001} and \citet{LeFeuvre:2008}.  The {\"O}pik-Wetherill method was used in both these studies, but they differred in their models of the Mars' crossing population.   \citet{Ivanov:2001} used the observed bright MCOs ($H<15$) while \citet{LeFeuvre:2008} used the debiased model of NEOs \citep{Bottke:2002} combined with the observed population of bright MCOs ($H<15$).  These source populations differ slightly from ours; we used the observed bright MCOs with $H<16$ as of June 2014.  Possibly the most important difference between our work and these previous studies is that they assumed uniform random distributions of the angular elements of MCOs.  Additionally, \citet{Ivanov:2001} neglected the gravitational focusing factor.   These previous studies do not quote a direct result for the current impact flux on Mars, but we can provide comparisons by making some minor calculations from their reported results. 

\citet{Ivanov:2001} provides the impact flux of the observed $H<18$ NEOs on the lunar surface, $0.77\times10^{-15} \rm{km}^{-2}  \rm{yr}^{-1}$, and writes that this is $40\%$ of the total impact flux (accounting for observational incompleteness of $H<18$ NEOs), which gives 
$F_{\!\!\rightmoon} (18) = 6.1 \times 10^{-23} \rm{km}^{-2} \rm{s}^{-1}$.  This is consistent with our estimate (in Section~\ref{s:ratio} of the lunar impact flux.  
Ivanov also tabulates the Mars/Moon impact flux ratio for three different values of Mars' eccentricity (Table~1 in his paper).  For the current value of Mars eccentricity of 0.093, his ratio is 4.8.  From this we obtain Ivanov's estimate for the current impact flux on Mars as $F(18) = 2.9 \times 10^{-22} \rm{km}^{-2}\rm{s}^{-1}$.

\citet{LeFeuvre:2008} report the lunar impact flux, $1.83 \times 10^{-15} \rm{km}^{-2} \rm{yr}^{-1}$, for NEOs larger than 1~km.  To convert to the flux of $H<18$ objects, we use the ratio of the number of NEOs brighter than $H=18$ and the number of NEOs larger than 1~km, 960/855, to calculate that \citet{LeFeuvre:2008}'s result for the lunar impact flux is equivalent to $F_{\!\!\rightmoon}(18)=6.5 \times 10^{-23} \rm{km}^{-2} \rm{s}^{-1}$. These authors also reported the current impact flux ratio of Mars/Moon as 3.23.  Thus, we obtain that their estimate of the current impact flux on Mars is equivalent to $F(18)=2.1 \times 10^{-22}\rm{km}^{-2} \rm{s}^{-1}$.

Our result, $F(18) = 3.0 \times 10^{-22} \rm{km}^{-2}\rm{s}^{-1}$, is approximately 50\% greater than \citet{LeFeuvre:2008}'s result, but nearly the same as \citet{Ivanov:2001}'s result.  The near-similarity of these results may appear surprising, because we expect our result to be smaller due to the fact that the non-uniform angular distribution depresses the impact flux, as we showed in Section~\ref{s:impactflux}.  Part of the explanation lies in a fortuitous near-cancellation of the effect of the non-uniform angular distribution (which reduces the flux) and the effect of gravitational focusing (which increases the flux), both of which were neglected in \citet{Ivanov:2001}'s calculation.  Another part owes to the different source populations.

\citet{Ivanov:2001} estimated the population of MCOs ($H<18$) by multiplying the population ratio of Earth-crossing objects (ECOs\footnote{\cite{Ivanov:2001} used NEOs as the terminology instead of ECOs.}), $N(H<18) / N(H<15)$, with the number of bright MCOs $N(H<15)$. From his estimate of 40\% observational completeness level of $H<18$ ECOs and his plot of the cumulative number distribution of observed ECOs, we estimate his ratio  $N(H<18) / N(H<15)\approx20$.  The number of bright ECOs and bright MCOs ($H<15$) in his work is 155 and 400, respectively, therefore we estimate the total number of MCOs ($H<18$) in his work is approximately eight thousand. In our model, the population ratio of MCOs $N(H<18) / N(H<15)$ is about 30 and the number of MCOs ($H<15$) is 516, which yields the estimated population of MCOs ($H<18$) of about fifteen thousand, about two times as large as Ivanov's. 

\citet{LeFeuvre:2008} did not report the number of MCOs ($H<18$) used in their calculations. It is likely  that their estimated population of MCOs $(H<18)$ is also much smaller than ours. Our $H$ magnitude  distribution for MCOs (Eq.~\ref{cumdist}) is steeper than that of NEOs in the range of $15<H<18$.  The population ratio of NEOs, $N(H<18) / N(H<15)$, is 19, which is smaller than our value for the population ratio of MCOs, $N(H<18) / N(H<15) \sim 30$.  Thus, their $\sim50\%$ smaller impact flux than ours is explained if they applied the former ratio to estimate the population of MCOs of $H<18$.

The current mean encounter velocities reported by \citet{Ivanov:2001} and \citet{LeFeuvre:2008} are 8.6 km/s and 9.1 km/s, respectively. These are slightly higher than our estimates of 8.5 km/s for the case of uniform angular distribution,  but lower than our result from the non-uniform angular distribution, 10.1 km/s. (Note that our estimates of mean encounter velocities are obtained for the bright MCOs, $H<16$, see section~\ref{s:impactvelocities}.)  The small difference in mean values of encounter velocities is understandable considering that the encounter velocity distribution has two strong peaks~(Fig.~\ref{ImpVel}).  We remark that the clearly double-peaked velocity distribution of impacts on Mars contradicts the model of a single-peak velocity distribution which has been widely used in Martian cratering studies~\citep[and references therein]{Flynn:1990,Williams:2014}.

%%%%%%%%%%%%%%%%%%%%%%%%%%%%%%%%%%%%%%%%%%%%%%%%%%%%%%%%%%%%%%%%%%%%%%%%%

\subsection{Variation of impact flux on secular timescales}\label{s:secular}

The impact flux estimates described above are specific to the present orbital configuration of Mars.  Over secular timescales, the mean impact flux evolves as Mars' orbit evolves, most importantly Mars' eccentricity vector relative to the mean eccentricity vector of MCOs (or relative to the orientation of Jupiter's perihelion direction).  Previous investigations of the impact flux on Mars averaged over secular timescales have considered only the secular variation of the magnitude of Mars' eccentricity but not its $\varpi$~\citep{Ivanov:2001,Ivanov:2002,LeFeuvre:2008}. However, as we demonstrated in Section~\ref{s:results}, the perihelion orientation is also important, just as much as the magnitude of its eccentricity. For example, when Mars' $\varpi$ is anti-aligned with Jupiter while the MCOs' eccentricity vectors tend to align with the $\varpi$ of Jupiter, the impact rate on Mars would be greater than the value calculated from the uniform angular distribution of MCOs. Therefore, it is important to account for the coupled eccentricity and apsidal variations to calculate the time--averaged impact flux on Mars. 

A full investigation of the effect of the non-uniform angular variables on the variation of the impact flux on Mars over secular timescales is beyond the scope of the present paper.  However, we make a few preliminary notes.  We carried out a numerical integration of the orbital evolution of the eight major planets (Mercury--Neptune) for $10^8$ years, with initial conditions at the present epoch.  We observe that the evolution of Mars' eccentricity and its longitude of perihelion relative to Jupiter's, $\varpi_{\mars}-\varpi_{\jupiter}$, are highly correlated; this is shown in Figure~\ref{RelLNP}.  The overall time-weighted distribution of $\varpi_{\mars}$ has a clear concentration near $\varpi_{\jupiter}$ (as indicated by the peak near zero of the solid line). Moreover, during the epochs when the eccentricity of Mars is high, exceeding 0.08, we observe that $\varpi_{\mars}$ is even more highly concentrated near Jupiter's and is rarely found anti-aligned with that planet.   The duration of epochs when Mars has high eccentricity (in excess of 0.08) account for 28\% of the whole $10^8$~yr simulation period.  The present-day Martian eccentricity, 0.093, lies within this high eccentricity range.  (We note that the probability of anti-alignment of Mars' perihelion relative to Jupiter's is vanishingly small when Mars' eccentricity exceeds $0.093$, i.e., its current value.)  \citet{Ivanov:2001} estimated that the time-averaged impact flux on Mars is less than $50\%$ of its present-day impact flux.  However, based on our investigation, we cautiously predict that the  long time averaged impact flux on Mars' may not be as strongly different from its current value as Ivanov claimed, because the current near-alignment of the Mars $\varpi$ with Jupiter's reduces the impact flux compared to the random angle distribution assumed in Ivanov's calculations.  Furthermore, the simulation results obtained by \citet{LeFeuvre:2008}, who also assumed random distributions of angular elements, find that the current impact flux is just 14\% higher than the long time average value.   Therefore, if the effect of the non-uniform anglular distribution of MCOs were included, we cannot rule out the possibility that the current impact flux is even lower than the time-averaged flux over secular timescales.

%------------------------------------------------------------------------------------------------------------------------------------------------------
\section{Summary and Conclusions}\label{s:conclusions}

In this work, we provide an updated and detailed calculation of the current impact flux on Mars, including its seasonal variation.  We also provide an updated  estimate for the Mars/Moon and Mars/Moon impact flux ratios.  We use a modification of the method of \citet{Opik:1951} and \citet{Wetherill:1967} for the collision probability of two independent Keplerian orbits.  Our impact flux calculation is based on the observationally nearly-complete population of bright Mars-crossing objects (respectively, Earth-crossing objects, for the lunar and terrestrial flux).  We pay careful attention to the non-uniform distribution of angular elements of this population and the effects of gravitational focusing.  We show that this non-uniform distribution, which is owed to the secular perturbations of the planets, strongly affects the mean impact flux on Mars as well as its seasonal variation.  It also affects the velocity distribution of impacts on Mars.  We also provide predictions for the meter-size impact flux on Mars, including its seasonal variation; we note that the magnitude of its seasonal variation is very sensitive to the orbital distribution, particularly the non-uniformity of the angular elements.  Observations of fresh Martian craters can test these predictions, and help to reveal the population and orbital distribution of the small size Mars-crossing objects.  A summary of our results is as follows.

\begin{enumerate}
\item
The population of Mars crossing objects has a strongly non-uniform distribution of the longitude of perihelion and the argument of perihelion, owing to secular planetary perturbations.  The longitude of perihelion has unidirectional distribution, approximately aligned with Jupiter's eccentricity vector.  The argument of perihelion has an axial distribution, with concentrations at $90^\circ$ and $270^\circ$.

\item
The current mean impact flux on Mars by minor planets brighter than $H=16$ is $2.9\times10^{-23}$~km$^{-2}$~s$^{-1}$.  If the non-uniform distribution of the angular elements of MCOs were to be neglected, this flux would be about twice as large.

\item
The impact flux of bright objects $(H<16)$ on Mars has a strong seasonal variation: it is at a maximum at Mars aphelion and minimum at its perihelion.  The impact flux $\pm30^\circ$ near aphelion is four times the impact flux near perihelion.  If the non-uniform distribution of the angular elements of MCOs were to be neglected, this ratio would be about four times as large.

\item
The impact velocity distribution has two distinct peaks, at 6.5~km~s$^{-1}$ and 11.5~km~s$^{-1}$, and a mean value of 11.5~km~s$^{-1}$.  It also has a strong seasonal variation, with a minimum at Mars' aphelion and a maximum at its perihelion.

\item
The Mars/Moon impact flux ratio is 5.0, and the Mars/Earth ratio is 2.8.  Our updated estimate for the current impact flux on the Moon and Earth by minor planets brighter than $H=18$ is $6\times10^{-23}$~km$^{-2}$~s$^{-1}$ and $1.0\times10^{-22}$~km$^{-2}$~s$^{-1}$, respectively.

\item
The flux of small impactors, $H<33$ (approximately meter-size) objects, on Mars is $(0.6-1.9)\times10^{-14}$~km$^{-2}$~s$^{-1}$, or roughly 100--300 impacts per year.  If the small impactors track the orbital distribution of the bright objects, then we estimate their impact rate near Mars' aphelion ($\pm30^\circ$) is four times that near perihelion ($\pm30^\circ$).  However if their angular elements have a uniform distribution (possibly due to Yarkovsky drift), then the aphelion-to-perihelion ratio would be as large as $\sim25$.   

\end{enumerate}

In the current era, we are in the stage of imaging fresh craters on Mars in almost real time with the help of orbiting space observatories, thus the seasonal variation of the impact rate can be available from crater counting \citep{Daubar:2012, Daubar:2013}.  As we have shown in this work, the seasonal variation is very sensitive to the perihelion distribution of the impactor population.  We therefore anticipate that the real angular distribution of small bodies and the importance of non-gravitational effects on their orbital distribution would be revealed from these observations. 

\acknowledgements
 This research was supported by NSF (grant AST-1312498) and NASA (grant NNX14AG93G), and made use of the NASA Astrophysics Data System Bibliographic Services and the Minor Planet Center database.

\bibliographystyle{icarus}
\bibliography{MarCrtSeas_v1.0}

\end{document}